\documentclass[aps,prd,amssymb,amsmath,nofootinbib]{revtex4}

\begin{document}

\title{Cosmological perturbations from stochastic gravity}
\author{Albert Roura}
\affiliation{Theoretical Division, T-8, Los Alamos National Laboratory,
M.S.~B285, Los Alamos, NM 87545}
\author{Enric Verdaguer}
\affiliation{Departament de F\'{\i}sica Fonamental and Institut de
Ci\`encies del Cosmos, Universitat de Barcelona,
Av.~Diagonal 647, 08028 Barcelona, Spain}

\begin{abstract}
In inflationary cosmological models driven by an inflaton field the
origin of the primordial inhomogeneities which are responsible for
large scale structure formation are the quantum fluctuations of the
inflaton field. These are usually computed using the standard theory
of cosmological perturbations, where both the gravitational and the
inflaton fields are linearly perturbed and quantized. The correlation
functions for the primordial metric fluctuations and their power
spectrum are then computed. Here we introduce an alternative procedure
for computing the metric correlations based on the Einstein-Langevin
equation which emerges in the framework of stochastic semiclassical
gravity.  We show that the correlation functions for the metric
perturbations that follow from the Einstein-Langevin formalism
coincide with those obtained with the usual quantization procedures
when the scalar field perturbations are linearized.  This method is
explicitly applied to a simple model of chaotic inflation consisting
of a Robertson-Walker background, which undergoes a quasi-de-Sitter
expansion, minimally coupled to a free massive quantum scalar
field. The technique based on the Einstein-Langevin equation can,
however, deal naturally with the perturbations of the scalar field
even beyond the linear approximation, as is actually required in
inflationary models which are not driven by an inflaton field such as
Starobinsky's trace-anomaly driven inflation or when calculating
corrections due to non-linear quantum effects in the usual inflaton
driven models.
\end{abstract}

%\pacs{}
%\date{\today}

\maketitle

\section{Introduction}

\label{sec1}

Inflation has become the paradigm for our understanding of the origin
of the primordial inhomogeneities which are responsible for large
scale cosmic structure. The typical inflationary scenario assumes a
period of accelerated expansion in the early universe, usually driven
by a scalar inflaton field, which provides a natural explanation for
the homogeneity, isotropy and flatness problems of the standard
big-bang cosmology
\cite{guth81,linde82a,albrecht82,linde83a,linde90}. The generation of
structure is explained by the back-reaction effect of the quantum
fluctuations of the inflaton field on the gravitational field which
translate, after quantization, into non-trivial two-point correlation
functions of the primordial gravitational fluctuations. These correlations
give an approximate Harrison-Zeldovich spectrum for large scales
\cite{mukhanov81,guth82,linde82b,hawking82,starobinsky82,bardeen83}.
The remarkable success of this scenario to explain the observed
anisotropies of the cosmic microwave background
\cite{smoot92,bennett03,peiris03,spergel07} is today the most
compelling reason which supports the inflationary paradigm
\cite{hollands02b,kofman02}, in spite of some interpretational
problems such as the transition from quantum to classical fluctuations
\cite{polarski96,kiefer98a,kiefer98b,kiefer98c,tanaka98,lombardo05,campo05,perez06}.

Semiclassical gravity is a mean field approximation that describes the
interaction of quantum matter fields with the gravitational field,
which is is treated as a classical geometry, and provides a suitable
framework for the study of macroscopic black holes as well as
scenarios in the early universe after the Planck time. In particular
it accommodates the different inflationary models. The key equation in
semiclassical gravity is the semiclassical Einstein equation where the
expectation value of the stress tensor operator of the quantum matter
fields is the source of the spacetime metric. In cosmology this is
usually assumed to be a spatially homogeneous and isotropic
Robertson-Walker spacetime. However, since this theory relies only on
the expectation value, it completely misses the fluctuations of the
stress tensor operator. Thus, when the back reaction of the
inhomogeneous fluctuations of the inflaton field around the
homogeneous background are relevant, as in the generation of
primordial inhomogeneities, the semiclassical equation is
insufficient.

In recent years a stochastic semiclassical gravity, or stochastic
gravity, approach has emerged as an extension of semiclassical gravity
which accounts for the quantum fluctuations of the stress tensor
\cite{hu03a,hu04a}. These fluctuations are characterized by the noise
kernel, which is defined as the symmetrized two-point quantum
correlation function of the stress tensor operator. The extension is
based on the so-called Einstein-Langevin equation, which is a
stochastic equation for the linearized gravitational perturbations
around a semiclassical background.  A Gaussian stochastic source with
a correlation function determined by the noise kernel is the key
ingredient of this equation. From the solutions of the
Einstein-Langevin equation, the two-point correlation functions for
the metric perturbations can be obtained.

Stochastic gravity provides an alternative framework to study the
generation of primordial inhomogeneities in inflationary models.
Besides the interest of the problem in its own right, there are also
other reasons that make this problem worth discussing from the point
of view of stochastic gravity. The Einstein-Langevin equation is not
restricted by the use of linearized perturbations of the inflaton
field. This may not be very important for inflationary models which
are driven by an inflaton field which takes a non-zero expectation
value, because the linear perturbations will give the leading
contribution; although the need to consider higher-order corrections
(from one-loop contributions) has recently been emphasized
\cite{weinberg05,weinberg06} (see also \cite{urakawa08b}). These
contributions are in any case important in models such as
Starobinsky's trace anomaly driven inflation \cite{starobinsky80},
which rely on conformally coupled scalar fields with a vanishing
expectation value. The corresponding Einstein equation is quadratic in
these fields and the linear approximation becomes trivial.

In this paper we prove that the usual quantization for linear
perturbations of both the metric and the inflaton field is equivalent
to using the Einstein-Langevin equation when the latter is restricted
to linearized inflaton perturbations, in the sense that the same
results for the relevant correlation functions of the metric
perturbations are obtained.

The plan of the paper is the following. In Sec.~II a brief description
of stochastic gravity is given. This is done in an axiomatic way by
showing that the semiclassical Einstein equation can be consistently
generalized in a perturbative way by including a Gaussian stochastic
source with vanishing expectation value defined through the noise
kernel. The dynamical equation for the metric perturbations is the
Einstein-Langevin equation.  An alternative derivation of this
equation, reviewed in Appendix~C, is based on the influence functional
method due to Feynman and Vernon, which is generally used to describe
the dynamics of open quantum systems, \emph{i.e.}, systems interacting
with an environment \cite{feynman63,feynman65}. Here the gravitational
field plays the role of the \emph{system} and the quantum matter
fields play the role of the \emph{environment}.

In Sec.~III we discuss the linearized perturbations around a
cosmological Robertson-Walker background coupled to a free massive
scalar field minimally coupled to the curvature. This corresponds to
the simplest model of chaotic inflation. The linearized
Einstein-Langevin equations is then used to obtain an expression for
the correlation function of the scalar-type metric perturbations. We
concentrate on metric perturbations of scalar type because they are
the only ones that couple to the inflaton perturbations in the linear
approximation. Next, we use the standard linear theory of cosmological
perturbations, quantize them and derive an expression for the
symmetrized quantum correlation function of the scalar metric
perturbations. This expression is then employed to show the
equivalence between this correlation function and that derived from
the Einstein-Langevin equation. An alternative proof of this
equivalence is provided in Appendix~\ref{appD}.

Note that whereas in stochastic gravity the metric is treated as a
classical but stochastic field, in the usual approach to linear
cosmological perturbations both metric and inflaton perturbations are
quantized. Nevertheless, the correlation functions derived within the
Einstein-Langevin approach agree with the symmetrized quantum
correlations, similarly to what happens in simpler open quantum
systems \cite{calzetta03a}. More specifically, the stochastic
correlation functions derived from the Einstein-Langevin equation
agree with the symmetrized quantum correlation functions of the theory
of gravity interacting with $N$ matter fields to leading order in
$1/N$ \cite{roura03b}; this was shown in Refs.~\cite{hu04b,hu04c} for
perturbations around a Minkowski background.  In this sense the
Einstein-Langevin equation can be regarded as a useful intermediary
tool to compute the quantum correlations of metric fluctuations in the
large $N$ approximation.  It should also be noted that there are
situations for open quantum systems where, for a sufficient degree of
environment-induced decoherence that guarantees the absence of
relevant interference effects, the temporal correlations of some
\emph{actual} properties of the system (corresponding to suitably
smeared projectors) can be described in terms of classical stochastic
processes governed by a Langevin equation \cite{gell-mann93}. In those
cases the stochastic correlation functions obtained from the Langevin
equation also describe such quasi-classical correlations of the system
dynamics.

It is, however, important to stress that when one linearizes with
respect to both the scalar metric perturbations and the inflaton
perturbations, as in the case discussed here, the system cannot be
regarded as a true open quantum system. The reason is that Fourier
modes decouple and the dynamical constraints due to diffeomorphism
invariance link the metric perturbations of scalar type with the
perturbations of the inflaton field so that only one true dynamical
degree of freedom is left for each Fourier mode.

In Sec.~IV we explicitly compute the correlation functions for the
scalar metric perturbations in a simple model of chaotic inflation
using the Einstein-Langevin approach as described in the previous
section.  A quasi de Sitter expansion for the background is assumed
and an almost Harrison-Zeldovich spectrum at large scales is
obtained. We comment on the different approximations that one is
naturally lead to consider within the two approaches.

Finally in Sec.~V we conclude by summarizing our results and briefly
discussing the changes that one would encounter when using the
Einstein-Langevin equation if the inflaton perturbations were treated
exactly, that is, beyond the linear approximation. We note in
particular that the scalar, vectorial and tensorial metric
perturbations are dynamically related in that case since they all
couple to the inflaton perturbations. The metric perturbations can
then be considered a true open quantum system.

Throughout the paper we use the $(+,+,+)$ convention of
Ref.~\cite{misner73}. We also make use of the abstract index notation
of Ref.~\cite{wald84}. The Latin indices ($a,b,c \ldots$) denote
abstract indices, whereas Greek indices are employed whenever a
particular coordinate system is considered [Latin indices such as
($i,j,k \ldots$) are used instead when referring only to spatial
components].

\section{Stochastic gravity formalism and Einstein-Langevin
equation}

\label{sec2}

\subsection{Stochastic gravity. General formalism}

\label{sec2.1}

There are a number of situations, especially in black hole physics and
cosmology, in which regarding spacetime as classical whereas the
remaining matter fields are quantized has proved very fruitful. This
is often considered a reasonable approximation as long as the typical
length scales involved are much larger than the Planck length. A first
step in that direction is to consider the evolution of quantum matter
fields on spacetimes with non-vanishing curvature. The consistent
formulation of quantum field theory on general globally hyperbolic
%\footnote{A given spacetime is said to be globally hyperbolic if there
%exists a Cauchy hypersurface, \emph{i.e.}, a hypersurface such that all the
%causal curves passing through any point of the spacetime cross the
%hypersurface exactly once.}
spacetimes is nowadays well established for free fields
\cite{birrell94,wald94}, and significant progress has been made for
interacting fields as well
\cite{hollands02a,hollands03,hollands07}. Up to this point the quantum
matter fields are regarded as test fields evolving on a fixed geometry
which is unaffected by their presence.

A second step is to consider the back reaction of quantum matter
fields on the spacetime geometry by including the expectation value of
the stress tensor operator of the quantum fields as a source of the
Einstein equation for the spacetime geometry, which becomes then the
so-called \emph{semiclassical Einstein equation}. The expectation
value of the stress tensor is divergent and a non-trivial
renormalization procedure is required even for a free field in order
to preserve general covariance. This can be achieved by using, for
instance, dimensional regularization or point splitting and
introducing suitable local counterterms, which are quadratic in the
curvature, in the bare gravitational action.  Any renormalization
method can be used provided Wald's axioms \cite{birrell94,wald94} are
satisfied, since this guarantees equivalent results.

The semiclassical Einstein equation was derived in
Ref.~\cite{hartle81} by considering the large $N$\/ limit of $N$\/
free scalar fields weakly interacting with the gravitational field so
that the product of the gravitational coupling constant times the
number of fields $N$\/ remains constant as $N$\/ tends to infinity;
see also Ref.~\cite{tomboulis77} for a related result concerning
fermions. However, the semiclassical Einstein equation is most often
introduced in an axiomatic way. The basic aspects of this framework,
commonly known as \emph{semiclassical gravity}
\cite{wald94,flanagan96}, can be summarized as follows. Let us
consider a manifold $\mathcal{M}$ with a Lorentzian metric $g_{ab}$
which is globally hyperbolic. Let us also consider a linear matter
field evolving on that manifold. In the Heisenberg picture the scalar
field operator $\hat{\phi}\left[ g\right] $ satisfies the Klein-Gordon
equation
\begin{equation}
\left( \nabla_a\nabla^a-m^{2}\right) \hat{\phi}(x)=0, \label{7.0}
\end{equation}
where $\nabla_a$ means covariant derivative with respect to the metric
$g_{ab}$, and the state of the scalar field, which is characterized by
a density matrix $\hat{\rho}\left[ g\right] $, is assumed to be
physically acceptable in the sense of Ref.~\cite{wald94}. This means
that it is of the so-called Hadamard type, so that the expectation
value for the stress tensor operator can be consistently renormalized.
The set $(\mathcal{M},g_{ab},\hat{\phi}\left[ g\right]
,\hat{\rho}\left[ g\right] )$ constitutes a self-consistent solution
of semiclassical gravity if the following semiclassical Einstein
equation is satisfied:
\begin{equation}
G_{ab}\left[ g\right] +\Lambda g_{ab}-2\left( \alpha A_{ab}\left[ g\right]
+\beta B_{ab}\left[ g\right] \right) =\kappa \langle \hat{T}_{ab}\left[
g\right] \rangle^\prime _\mathrm{ren}\text{,}  \label{7.1}
\end{equation}
where $G_{ab}$ is the Einstein tensor, $\langle \hat{T}_{ab}\left[
  g\right] \rangle^\prime _\mathrm{ren}$ is the suitably renormalized
expectation value of the stress tensor operator corresponding to the
scalar field operator $\hat{\phi}\left[ g\right] $ and $\alpha $,
$\beta $, $\Lambda $ and $\kappa$ are renormalized parameters (the
prime in the expectation value is used to distinguish it from the
expectation value introduced below). We considered natural units in
which $\hbar =c=1$, and introduced the notation $\kappa = 8 \pi G=8\pi
/m_{p}^{2}$ for the renormalized gravitational coupling constant where
$m_{p}$ is the Planck mass.  The local tensors $A_{ab}$ and $B_{ab}$
are obtained by functionally differentiating with respect to the
metric terms in the action that correspond to the Lagrangian densities
proportional to $C^{abcd} C_{abcd}$ and $R^{2}$, respectively, where
$C_{abcd}$ and $R$ are the Weyl tensor and the scalar curvature. These
terms
%are closely related to
correspond to the finite part of the counterterms introduced in the bare
gravitational action to cancel the divergences arising in the
expectation value of the stress tensor \cite{birrell94}.

{}From now on, and despite their purely geometric character, we will
consider for notational simplicity that the last three terms on the
left-hand side of Eq.~(\ref{7.1}) have been reabsorbed in the
renormalized expectation value of the stress tensor operator, which we
write now without the prime; this can be done consistently because
$\nabla^a A_{ab}=0=\nabla^a B_{ab}$. Taking this into account, the
semiclassical Einstein equation becomes
\begin{equation}
G_{ab}\left[ g\right] =\kappa \langle \hat{T}_{ab}\left[ g\right]
\rangle _\mathrm{ren}\text{,}  \label{7.2}
\end{equation}
where we should now keep in mind that the expectation value depends on
the renormalized parameters $\Lambda$, $\alpha$ and $\beta$.

There are, however, situations in which the fluctuations of the stress
tensor operator are important \cite{ford82,kuo93,phillips97}. In
Refs.~\cite {martin99a,martin99b} it was shown that the semiclassical
Einstein equation (\ref{7.1}) could be consistently extended to
partially account for the fluctuations of the stress tensor operator
by introducing a Gaussian stochastic source. More precisely, given a
self-consistent solution of semiclassical gravity one can introduce
the following equation for the metric perturbations $h_{ab}$ around
the background metric $g_{ab}$:
\begin{equation}
G_{ab}\left[ g+h\right] +\Lambda \left( g_{ab}+h_{ab}\right) -2\left( \alpha
A_{ab}\left[ g+h\right] +\beta B_{ab}\left[ g+h\right] \right) =\kappa
\langle \hat{T}_{ab}\left[ g+h\right] \rangle^\prime _\mathrm{ren}+\kappa \,\xi
_{ab}\left[ g\right] \text{,}  \label{7.3}
\end{equation}
where the whole equation should be understood to linear order in
$h_{ab}$. Note that throughout this paper indices will be raised and
lowered using the background metric. The renormalized expectation
value is computed with the scalar field operator satisfying the
Klein-Gordon equation on the perturbed metric $g_{ab}+h_{ab}$ and the
Gaussian stochastic source $\xi _{ab}$ is completely determined by the
following correlation functions:
\begin{eqnarray}
\left\langle \xi _{ab}\left[ g;x\right) \right\rangle _{\xi } &=&0\text{,}
\label{7.4} \\
\left\langle \xi _{ab}\left[ g;x\right) \xi _{cd}\left[ g;y\right)
\right\rangle _{\xi } &=&N_{abcd}\left( x,y\right) =\frac{1}{2}\left\langle
\left\{ \hat{t}_{ab}\left[ g;x\right) ,\hat{t}_{cd}\left[ g;y\right) \right\}
\right\rangle \text{,}  \label{7.5}
\end{eqnarray}
where we used $\left\langle \ldots \right\rangle _{\xi }$ to denote
the expectation value with respect to the stochastic classical source
$\xi _{ab}\left[ g\right] $. The operator $\hat{t}_{ab}\left[ g\right]
$ is defined as $\hat{t}_{ab}\left[ g\right] \equiv \hat{T}_{ab}\left[
  g\right] -\langle \hat{T}_{ab}\left[ g\right] \rangle $ and the
bitensor $N_{abcd}\left( x,y\right)$, which determines the correlation
function of the stochastic source, is computed using the scalar field
operator satisfying the Klein-Gordon equation for the background
metric $g_{ab}$.  The bitensor
$N_{abcd}(x,y)$ is called the \emph{noise kernel}, it describes the
quantum fluctuations of the stress tensor operator and is
positive-semidefinite.  Strictly speaking, the previous definition for
the operator $\hat{t}_{ab}$ only makes sense when some kind of
regulator is employed since both the operator $\hat{T}_{ab}\left[
  g\right] $ and the expectation value $\langle \hat{T}_{ab}\left[
  g\right] \rangle $ are divergent.  However, the operator
$\hat{t}_{ab}$ is finite in the sense that one can compute any matrix
element of this operator using a regularized version of the two terms
that define it, and one finally gets a finite result when removing the
regulator because the divergences coming from both terms cancel out
exactly \cite{martin99b}. Hence, the noise kernel requires no renormalization whereas
the divergences of the expectation value $\langle \hat{T}_{ab}\left[
  g+h\right] \rangle $ appearing in Eq.~(\ref{7.3}) are canceled by
the counterterms whose finite contribution corresponds to the last
three terms on the left-hand side. Furthermore, since $\left\langle
  \xi _{ab}\left[ g\right] \right\rangle _{\xi }=0$, Eq.~(\ref{7.3}),
which is called the \emph{Einstein-Langevin equation}, reduces to the
semiclassical Einstein equation for the metric perturbations $h_{ab}$
around the background metric $g_{ab}$ when taking the expectation
value with respect to the stochastic source $\xi _{ab}$.%
\footnote{Remember that Eq.~(\ref{7.3}) should be understood to linear
order in $h_{ab}$.
%Since the background metric $g_{ab}$ satisfies Eq.~(\ref{7.1}), one is
%only left with terms linearly proportional to $h_{ab}$ in addition to the
%stochastic source.
Therefore, when taking the average over all possible realizations of the stochastic source, the equation satisfied by $\langle h_{ab} \rangle_\xi$ coincides with that obtained by linearly perturbing the semiclassical Einstein equation~(\ref{7.1}).}
This framework, in which the metric perturbations are regarded as a
stochastic process satisfying the Einstein-Langevin equation, is
usually referred to as \emph{stochastic gravity}.

Similarly to what was done for the semiclassical Einstein equation, we
will assume that the last three terms on the left-hand side of
Eq.~(\ref{7.3}) are reabsorbed in the renormalized expectation value
of the stress tensor operator, so that the Einstein-Langevin equation
will be written from now on as
\begin{equation}
G_{ab}^{(1)}\left[ g+h\right] =\kappa \langle \hat{T}_{ab}^{(1)}\left[
g+h\right] \rangle _\mathrm{ren}+\kappa \,\xi _{ab}\left[ g\right] \text{,}
\label{7.7}
\end{equation}
where the superindex $(1)$ means that only terms linear in the metric
perturbations $h_{ab}$ are kept. This follows straightforwardly from
the fact that Eq.~(\ref{7.3}) was considered only to linear order in
$h_{ab}$ (the stochastic source $\xi _{ab}$ is regarded to be of the
same order as $h_{ab}$) and that the zero order contribution is
identically satisfied, since the background configuration was assumed
to be a solution of semiclassical gravity.

A necessary condition for the integrability of the Einstein-Langevin
equation, via the Bianchi identity, is the conservation of the
stochastic source.  Hence, one must make sure that the stochastic
source $\xi_{ab}\left[g\right]$ is covariantly conserved so that
Eq.~(\ref{7.7}) is a consistent extension of the semiclassical
Einstein equation (\ref{7.2}). That the stochastic process $\nabla
^{a}\xi _{ab}(x)$ vanishes is a consequence of the stress tensor
conservation on the background metric \cite{martin99a,martin99b}.
Furthermore, it can also be checked that the Einstein-Langevin
equation is compatible with the gauge symmetry corresponding to
infinitesimal diffeomorphisms. In fact, both the stochastic source and
the remaining terms of the Einstein-Langevin equation are separately
invariant under gauge transformations for the metric perturbations of
the form $h_{ab}\rightarrow h_{ab}+\nabla _{a}\zeta _{b}+\nabla
_{b}\zeta _{a}$ corresponding to infinitesimal diffeomorphisms
generated by any arbitrary vector field $\vec{\zeta}$ defined on the
background spacetime \cite{martin99a,martin99b}.

%Finally, it is worth making a remark on the contribution of the stochastic
%source to the trace anomaly. The classical stress tensor for a massless
%scalar field conformally coupled to the spacetime curvature is traceless, as
%follows from the invariance of the classical action under conformal
%transformations in that case, and the same is true for a massless free
%fermionic field). However, the counterterms introduced in order to
%renormalize the expectation value of the stress tensor quantum operator break
%such an invariance and actually give a non-vanishing contribution of purely
%geometrical character to the trace of the expectation value of the stress
%tensor operator \cite{birrell94}. This contribution, which is known as the
%trace anomaly, can also be interpreted to yield a contribution which is
%proportional to the identity operator to the trace of the suitably
%regularized stress tensor operator. This implies
%$\hat{t}_{a}^{a}=0$ and, consequently, $\left\langle \xi _{a}^{a}(x)\xi
%_{c}^{d}(y)\right\rangle_{\xi }=\frac{1}{2}\left\langle \left\{ \hat{t}_{a}
%^{a}(x),\hat{t}_{c}^{d}(y)\right\} \right\rangle = 0$. Thus, the stochastic
%source gives no contribution to the trace anomaly for a conformally coupled
%massless scalar field.

We finish this general introduction to the Einstein-Langevin equation
by briefly mentioning that there are derivations of the
Einstein-Langevin equation in different cosmological settings making
use of functional methods
\cite{calzetta94,hu95a,hu95b,campos96,calzetta97c} or a derivation
using arguments based on the renormalization group
\cite{lombardo97}. In Appendix~\ref{appB} we sketch a derivation of
the Einstein-Langevin equation~(\ref{7.7}) for the case of a general
globally hyperbolic background spacetime using the influence
functional formalism \cite{martin99b,martin99c}.
The Einstein-Langevin equation has also been applied to the study of
fluctuations in black hole spacetimes \cite{hu07a,roura07a,hu07b}.

\subsection{Einstein-Langevin equation for cosmological perturbations}

\label{sec2.2}

In this paper we will study small perturbations around a
Robertson-Walker background when the matter source is a
minimally-coupled scalar field with a quadratic potential.  In fact,
this corresponds to the simplest model of chaotic inflation with the
scalar field playing the role of the inflaton field, but it is
sufficient for our purpose of illustrating the relationship between
the usual treatment of cosmological perturbations and those approaches
based on the Einstein-Langevin equation within the framework of
stochastic gravity. Furthermore, taking into account the assumptions
made throughout the forthcoming sections, the generalization of our
main conclusions and results to non-linear potentials should be rather
straightforward as long as we keep to quadratic order in the scalar
field perturbations when considering the potential.

Recall that the form for the line element of a general Robertson-Walker
metric is
\begin{equation}
ds^{2}=-dt^{2}+a^{2}\left( t\right) \gamma _{ij}dx^{i}dx^{j}\text{,}
\label{7.101}
\end{equation}
where $a\left( t\right) $ is called the scale factor and $\gamma
_{ij}$ is the induced metric for the homogeneous spatial sections,
which are maximally symmetric hypersurfaces. The line element of the
spatial sections can have the three following forms $\gamma_{ij}dx^i
dx^j = \{d\chi ^{2}+\sin ^{2}\chi d\Omega ^{2},\ dr^{2}+r^{2}d\Omega
^{2},\ d\chi ^{2}+\sinh ^{2}\chi d\Omega ^{2} \}$ depending on whether
the curvature is positive, zero, or negative, respectively.  In terms
of the conformal time coordinate $\eta =\int dta^{-1}(t)$ the metric
(\ref{7.101}) becomes
\begin{equation}
ds^{2}=a^{2}\left( \eta \right) \left( -d\eta ^{2} +
\gamma _{ij}dx^{i}dx^{j}\right) \text{.}  \label{7.103}
\end{equation}

Before proceeding further it is convenient to introduce the following
decomposition for the scalar inflaton field, which will be used
throughout:
\begin{equation}
\hat{\phi}(x)=\phi (\eta )+\hat{\varphi}(x)\text{,}  \label{8.0}
\end{equation}
where $\phi (\eta )$, which corresponds to the expectation value
$\langle \hat{\phi}[g;x)\rangle$ of the inflaton field on the
background metric, is a homogeneous classical-like (as an operator it
is proportional to the identity) solution of the Klein-Gordon equation
which is compatible with the background metric through the
semiclassical Einstein equation (\ref{7.2}).  The operator
$\hat{\varphi}(x)$, which will be referred to as the \emph{inflaton
  field perturbations}, corresponds to the quantum operator for a
minimally-coupled massive scalar field whose expectation value
vanishes on the background spacetime, \emph{i.e.}, $\langle
\hat{\varphi}[g;x)\rangle =0$. We will consider a Gaussian state for
the inflaton field and, thus, for the inflaton field perturbations;
see Appendix~\ref{app0} for the definition and the basic properties of
pure Gaussian states and the relationship between the state of the
inflaton field and the inflaton field perturbations.

It should also be stressed that there are many situations
(\emph{e.g.}, in the context of stochastic inflation) in which the
classical background configuration of the inflaton field will not be
homogeneous over the whole spacetime.  Nevertheless, this will not
have observable consequences at present provided that the scale of the
inhomogeneities is larger than the horizon before the last $60$
$e$-folds of inflation.
% ********* comment on stochastic and chaotic inflation **********
In fact, when studying models of eternal inflation
\cite{vilenkin83b,linde86,linde94} using the formalism of stochastic
inflation \cite{starobinsky86}, the expectation value of the inflaton
field is no longer the relevant object. One should consider instead
the amplitude of a given realization of the inflaton field %operator
smeared over scales slightly larger than the horizon radius right
before the region that had left the self-regenerating regime and would
eventually give rise to our visible universe underwent the last $60$
$e$-folds of inflation. It has been argued that in those circumstances
the smeared inflaton field behaves as a classical stochastic
process. (This is closely related to the quantum to classical
transition problem for the inflaton fluctuations
\cite{polarski96,kiefer98a,kiefer98b,kiefer98c,tanaka98,lombardo05,campo05,perez06}.)
If that is the case, one can use a particular realization of the
smeared inflaton right before the last $60$ $e$-folds of inflation as
the classical background configuration $\phi (\eta )$ and treat it in
the same way in which one would have dealt with a quantum expectation
value.

Let us begin by discussing the semiclassical Einstein equation
(\ref{7.2}) for the background metric $g_{ab}$ defined by
Eq.~(\ref{7.101}).  The right-hand side of Eq.~(\ref{7.2}) is the
properly renormalized expectation value for the stress tensor of the
inflaton field operator, which satisfies the Klein-Gordon
equation~(\ref{7.0}) on the background spacetime.

If we consider the general expression for the stress tensor operator
of a minimally-coupled massive scalar field
\begin{equation}
\hat{T}_{ab}=\nabla _{a}\hat{\phi}\nabla _{b}\hat{\phi}-\frac{1}{2}%
g_{ab}(g^{cd}\nabla _{c}\hat{\phi}\nabla _{d}\hat{\phi}+m^{2}\hat{\phi}^{2})%
\text{,}  \label{8.3}
\end{equation}
and use the decomposition of the scalar field introduced in
Eq.~({\ref{8.0}}), the expectation value for the stress tensor
operator can be separated into three different contributions:
\begin{equation}
\langle \hat{T}_{ab}[g]\rangle _\mathrm{ren}=\langle \hat{T}_{ab}[g]\rangle _{\phi
\phi }+\langle \hat{T}_{ab}[g]\rangle _{\phi \varphi }+\langle \hat{T}%
_{ab}[g]\rangle _{\varphi \varphi }^\mathrm{ren}\text{,}  \label{8.4}
\end{equation}
where the subindices $\phi\phi$, $\phi\varphi$ and $\varphi\varphi$
are used to denote the contributions to the stress tensor operator
which are respectively quadratic in $\phi(\eta)$, linear in both
$\phi(\eta)$ and $\hat{\varphi}(x)$, and quadratic in
$\hat{\varphi}(x)$.  The first term depends just on the homogeneous
solution $\phi (\eta)$, the second term vanishes since it is
proportional to $\langle \hat{\varphi}[g;x)\rangle$ and the third
term, which is completely independent of the homogeneous part $\phi
(\eta )$, is quadratic in the inflaton field perturbations
$\hat{\varphi}(x)$ and needs renormalization.

The first term on the right-hand side of Eq.~(\ref{8.4}) will be
denoted by $\mathcal{T}_{ab}\equiv\langle \hat{T}_{ab}[g]\rangle
_{\phi \phi }$; see Appendix~\ref{appA} for further comments on this
notation. Taking into account the special form of the Robertson-Walker
metric, in the basis associated with the conformal time and comoving
spatial coordinates these components can be rewritten as
\begin{eqnarray}
\mathcal{T}_{00} &=&\frac{1}{2}\left( \left( \phi ^{\prime }\right)
^{2}+m^{2}a^{2}\phi ^{2}\right) \text{,}  \label{8.5} \\
\mathcal{T}_{ij} &=&\frac{1}{2}\left( \left( \phi ^{\prime }\right)
^{2}-m^{2}a^{2}\phi ^{2}\right) \gamma _{ij}\text{,}  \label{8.6}
\end{eqnarray}
where primes denote derivatives with respect to the conformal time
$\eta $. In this coordinate system the time-time and space-space
components of $a^{-2}(\eta)\mathcal{T}_{\mu\nu}$ can be respectively
identified with the energy density $\rho (\eta)$ and the isotropic
pressure $p(\eta)$ of a perfect fluid. The components of
Eq.~(\ref{7.2}) become then the usual Friedmann equations
\begin{eqnarray}
\frac{\kappa }{2}\left( \left( \phi ^{\prime }\right) ^{2}+m^{2}a^{2}\phi
^{2}\right)  &=&3\left( \mathcal{H}^{2}+\epsilon \right) \text{,}
\label{8.8} \\
\frac{\kappa }{2}\left( \left( \phi ^{\prime }\right) ^{2}-m^{2}a^{2}\phi
^{2}\right)  &=&-\left( 2\mathcal{H}^{\prime }+\mathcal{H}%
^{2}+\epsilon \right) \text{,}  \label{8.9}
\end{eqnarray}
%\begin{eqnarray}
%\frac{\kappa }{2}\left( \left( \phi ^{\prime }\right) ^{2}+m^{2}a^{2}\phi
%^{2}\right)  &=&3a^{-2}\left( \mathcal{H}^{2}+\epsilon \right) \text{,}
%\label{8.8} \\
%\frac{\kappa }{2}\left( \left( \phi ^{\prime }\right) ^{2}-m^{2}a^{2}\phi
%^{2}\right)  &=&-a^{-2}\left( 2\mathcal{H}^{\prime }+\mathcal{H}%
%^{2}+\epsilon \right) \text{,}  \label{8.9}
%\end{eqnarray}
where $\mathcal{H}=a^{\prime }/a$ and $\epsilon =0,1,-1$ depending on
whether the homogeneous spatial sections of the Robertson-Walker
geometry are respectively flat, with positive curvature or with
negative curvature.

The third term on the right-hand side of Eq.~(\ref{8.4}), $\langle
\hat{T}_{ab}[g]\rangle _{\varphi \varphi }^\mathrm{ren}$, will in turn
have a similar structure to that of Eqs.~(\ref{8.5}) and (\ref{8.6})
with diagonal non-vanishing components which can be regarded as
corrections $\Delta \rho (\eta )$ and $\Delta p(\eta )$ to the energy
density and pressure. This structure is necessary so that the
solutions of Eq.~(\ref{7.2}) are of Robertson-Walker type, but there is a family of quantum states of the scalar field which gives rise to such a structure for $\langle
\hat{T}_{ab}[g]\rangle _{\varphi \varphi }^\mathrm{ren}$. They can be characterized as follows. Since the Lie derivatives of the six spacelike
Killing vectors which characterize a Robertson-Walker metric commute
with the Klein-Gordon operator satisfying Eq.~(\ref{7.0}), one
can introduce a unitary operator which implements at the quantum level
the symmetries corresponding to the six Killing vectors and is
preserved by the dynamical evolution. Consequently, the Hadamard
function (the quantum expectation value of the anticommutator of the
field) employed to compute the renormalized expectation value of the
field $\hat{\varphi}(x)$ will respect the symmetries of the
Robertson-Walker geometry provided that one considers a quantum
initial state which is kept invariant, up to a phase, by the unitary
operator associated with those symmetries. Throughout this paper we
will consider this class of states (spatially homogeneous and isotropic).
Nevertheless, being quadratic in the inflaton
perturbations, which are considered in general to be much smaller
during the inflationary period than the homogeneous background
solution $\phi (\eta )$, the contribution from the last term in
Eq.~(\ref{8.4}) and, hence, the corrections $\Delta \rho $ and $\Delta
p$, will in general be small compared to those from Eqs.~(\ref{8.5})
and (\ref{8.6}) during the inflationary period. The usual treatments
which keep to linear order in both the metric perturbations and the
inflaton perturbations directly discard them.  This is actually the
situation that we will be interested in here.  Therefore, the
background solution for the scale factor $a(\eta)$ is completely
determined by Eqs.~(\ref{8.8}) and (\ref{8.9}) without considering the
corrections that come from the third term on the right-hand side of
Eq.~(\ref{8.4}), which is approximated to linear order by $\langle
\hat{T}_{ab}[g]\rangle_\mathrm{ren} \approx \mathcal{T}_{ab}.$

In addition, either from the conservation of the stress tensor,
$\nabla^{a}\mathcal{T}_{ab}=0$, or by taking the expectation value of
Eq.~(\ref{7.0}), the homogeneous background solution $\phi (\eta )$ is
seen to satisfy the following Klein-Gordon equation on the background
Robertson-Walker metric:
\begin{equation}
\phi ^{\prime \prime }+2\frac{a^{\prime }}{a}\phi ^{\prime }+ma^{2}\phi =0
\text{.}  \label{8.10}
\end{equation}

Let us now consider the objects which appear in the Einstein-Langevin
equation (\ref{7.7}) and particularize them to the case addressed
here. The geometric part,\emph{\ i.e.}, the components of the Einstein
tensor for a linear perturbation $h_{ab}$ of the metric will be
discussed in the next section. The contribution to the expectation
value of the stress tensor which is linear in the metric perturbation,
$\langle \hat{T}_{ab}^{(1)}[g+h]\rangle _\mathrm{ren}$, can be
decomposed according to Eq.~(\ref{8.0}) as:
\begin{equation}
\langle \hat{T}_{ab}^{(1)}[g+h]\rangle _\mathrm{ren}=\langle
\hat{T}%
_{ab}^{(1)}[g+h]\rangle _{\phi \phi }+\langle \hat{T}%
_{ab}^{(1)}[g+h]\rangle _{\phi \varphi }+\langle \hat{T}%
_{ab}^{(1)}[g+h]\rangle _{\varphi \varphi }^\mathrm{ren}\text{,}  \label{8.21}
\end{equation}
where the whole inflaton field satisfies now the Klein-Gordon equation
on the perturbed metric $\tilde{g}_{ab}=g_{ab}+h_{ab}$,
\begin{equation}
( \tilde\nabla_a\tilde\nabla^a-m^{2}) \hat{\phi}=0,
\label{8.21b}
\end{equation}
and $\tilde\nabla_a$ means the covariant derivative with respect to
$\tilde{g}_{ab}$.  The first term on the right-hand side of
Eq.~(\ref{8.21}), $\langle \hat{T} _{ab}^{(1)}\rangle _{\phi \phi }$,
depends on the scalar field only via the homogeneous background
solution $\phi (\eta )$, which was already fixed by Eq.~(\ref{8.10})
together with Eqs.~(\ref{8.8}) and (\ref{8.9}), and therefore the
metric perturbations enter only through the explicit dependence of the
stress tensor on the metric. Contrary to what happened in
Eq.~(\ref{8.4}), the second term on the right-hand side of
Eq.~(\ref{8.21}), $\langle \hat{T} _{ab}^{(1)}\rangle _{\phi \varphi
}$, no longer vanishes since it is now proportional to $\left\langle
  \hat{\varphi}[g+h]\right\rangle $ and the Klein-Gordon equation
satisfied by $\hat{\varphi}[g+h]$ on the spacetime with the perturbed
metric $\tilde{g}_{ab}$, given by Eq.~(\ref{8.21b}), has an
inhomogeneous source term proportional to the metric perturbation
$h_{ab}$ and the homogeneous background solution $\phi (\eta )$ which
in general prevents its expectation value $\left\langle
  \hat{\varphi}[g+h]\right\rangle $ from vanishing. Hence, the only
non-vanishing contributions to the second term are those which depend
implicitly on the metric perturbations through the quantum operator
for the inflaton perturbations $\hat{\varphi} [g+h]$.

Finally, the third term, $\langle \hat{T} _{ab}^{(1)}\rangle _{\varphi
  \varphi }^\mathrm{ren}$, which requires renormalization, will have
contributions with either explicit or implicit dependence on the
metric perturbation. Not only the contributions which depend on the
metric perturbations explicitly, but also those which depend
implicitly via $\hat{\varphi}[g+h]$ are ultimately quadratic in the
inflaton perturbations $\hat{\varphi}[g]$ [after solving
Eq.~(\ref{8.21b}) perturbatively in the metric perturbations];
otherwise they would vanish, as follows from the fact that
$\left\langle \hat{\varphi}[g]\right\rangle =0$.

Similarly to what was said concerning the last term in
Eq.~(\ref{8.4}), the last term in Eq.~(\ref{8.21}) is not taken into
account by usual approaches to cosmological perturbations, which keep
to linear order in the inflaton perturbations as well as the metric
perturbations. We will not consider these terms either in the next two
sections, but some general remarks on how to deal with them, and
possible implications, will be made in Sec.~\ref{sec5}.

Let us now briefly concentrate on the noise kernel, which accounts for
the stress tensor fluctuations and characterizes the correlations of
the stochastic source $\xi _{ab}$. It is proportional to $\left\langle
  \left\{ \hat{t}_{ab}[g],\hat{t}_{cd}[g]\right\} \right\rangle$ where
$\hat{t}_{ab}=\hat{T}_{ab}-\langle \hat{T}_{ab}\rangle$, and is
evaluated on the background metric. Using Eqs.~(\ref{8.0}) and
(\ref{8.3}) it can be separated into the following non-vanishing
terms:
\begin{equation}
\left\langle \left\{ \hat{t}_{ab}[g],\hat{t}_{cd}[g]\right\} \right\rangle
=\left\langle \left\{ \hat{t}_{ab}[g],\hat{t}_{cd}[g]\right\} \right\rangle
_{\phi \varphi }+\left\langle \left\{ \hat{t}_{ab}[g],\hat{t}_{cd}[g]\right\}
\right\rangle _{\varphi \varphi }\text{,}  \label{8.22}
\end{equation}
where the first and second terms on the right-hand side are,
respectively, quadratic and quartic in the inflaton perturbation
$\hat{\varphi}[g]$.  We used a notation similar to that introduced in
Eq.~(\ref{8.4}) since the first term on the right-hand side of
Eq.~(\ref{8.22}) comes entirely from those contributions to the
operators $\hat{t}_{ab}$ and $\hat{t}_{cd}$ which are proportional to
both $\phi(\eta)$ and $\hat{\varphi} [g]$, whereas the last term in
Eq.~(\ref{8.22}) comes from the contributions to the stress tensor
which are quadratic in $\hat{\varphi}[g]$.  The contribution to the
noise kernel which depends on the background homogeneous solution
$\phi(\eta)$ but not on the inflaton perturbation $\hat{\varphi}$
vanishes since, being proportional to the identity, the corresponding
stress tensor operator coincides with its expectation value.
%and, thus, $\left(\hat{t}_{ab}\right)_{_{\phi \phi }}=\left(\hat{T}_{ab}
%\right)_{\phi\phi}-\left\langle \hat{T}_{ab}\right\rangle_{\phi\phi}=0$.
The terms linear in $\hat{\varphi}[g]$ also vanish because
$\left\langle \hat{\varphi}[g]\right\rangle =0$. Finally, since we
will be considering Gaussian quantum states for the inflaton
perturbations (see Appendix~\ref{app0} for a definition and a brief
description of some basic properties of Gaussian states), the
$3$-point quantum correlation functions $\left\langle
  \hat{\varphi}[g]\hat{\varphi}[g]\hat{\varphi}[g]\right\rangle $ are
proportional to the expectation value $\left\langle \hat{\varphi}[g]
\right\rangle$ and, therefore, the contributions which are cubic in
$\hat{\varphi}[g]$ vanish as well.

It is important to note that both the quadratic and the quartic
contributions to the noise kernel are separately conserved since both
$\phi (\eta)$ and $\hat{\varphi}[g;x)$ independently satisfy the
Klein-Gordon equation (\ref{8.0}) on the background geometry; recall
that $\phi(\eta) = \langle \hat{\phi}[g]\rangle$. Due to this fact, we
can consistently consider a pair of independent stochastic sources
$\xi _{1\,ab}$ and $\xi _{2\,ab}$ associated with each term so that
$\xi_{ab} =\xi _{1\,ab}+\xi _{2\,ab}$ with
$\left\langle \xi _{1\,ab}(x)\xi _{1\,cd}(x^{\prime })\right\rangle
=\left\langle \left\{ \hat{t}_{ab}(x),\hat{t}_{cd}(x^{\prime })\right\}
\right\rangle _{\phi \varphi }[g]$, $\left\langle \xi _{2\,ab}(x)\xi
_{2\,cd}(x^{\prime })\right\rangle =\left\langle \left\{ \hat{t}_{ab}(x),
\hat{t}_{cd}(x^{\prime })\right\} \right\rangle _{\varphi \varphi }[g]$ and
$\left\langle \xi _{1\,ab}(x)\xi _{2\,cd}(x^{\prime })\right\rangle =0$. The
integrability of the Einstein-Langevin equation with any of the two sources
is then guaranteed because both sources are separately conserved.

In the next section we will show that keeping only $\xi _{1\,ab}$,
which can be thought to be of the same order as $\hat{\varphi}$, the
results obtained using the Einstein-Langevin equation and those from
the usual treatments which quantize the linearized theory for both the
metric and the inflaton perturbations are equivalent. On the other
hand, some of the main features and consequences of the source $\xi
_{2\,ab}$, which can be regarded as being of quadratic order in
$\hat{\varphi}$, will be briefly discussed in Sec.~\ref{sec5} and
studied in more detail in Ref.~\cite{roura03b}.
Of course, when considering the stochastic source $\xi _{2\,ab}$,
the last term on the right-hand side of Eqs.~(\ref{8.4}) and (\ref{8.21})
should also be considered since their contribution is of the same
order as that of $\xi _{2\,ab}$.

\section{Einstein-Langevin equation for linearized cosmological
perturbations}

\label{sec3}

\subsection{Gauge invariant formalism for linearized cosmological
perturbations}

\label{sec3.1}

Let us consider small metric perturbations around a fixed
Robertson-Walker background geometry. It can be shown \cite{stewart90}
that the most general expression for the components of metric
perturbations in some particular coordinate system can be written as
%\begin{equation}
%h_{\mu \nu }=a^{2}(\eta )\left(
%\begin{array}{cc}
%2\phi & -\left( B_{|i}+S_{i}\right) \\
%-\left( B_{|i}+S_{i}\right) & 2\left( \psi \gamma _{ij}-E_{|ij}\right)
%+\left( F_{i|j}+F_{j|i}\right) +h_{ij}
%\end{array}
%\right) \text{,}  \label{8.20}
%\end{equation}
\begin{eqnarray}
h_{00}&=& 2\bar\phi \; a^2(\eta),\nonumber\\
h_{0i}&=& -\left( B_{|i}+S_{i}\right)a^{2}(\eta ),\nonumber\\
h_{ij}&=& \left[2\left( \psi \gamma _{ij}-E_{|ij}\right)
          +\left( F_{i|j}+F_{j|i}\right) +h_{ij}\right]a^2(\eta),
\label{8.20}
\end{eqnarray}
which depends on ten functions: $\bar\phi $, $\psi $, $B$ and $E$, the
two independent components of each transverse vector $S_{i}$ and
$F_{i}$, and the two independent components of the traceless and
transverse symmetric tensor $h_{ij}$.  The vectors $S_{i}$ and $F_{i}$
as well as the tensor $h_{ij}$ are tangent to the isotropic and
homogeneous spatial sections of the Robertson-Walker spacetime, but
depend in general on the conformal time $\eta $ which labels each
spatial section.  Furthermore, the notation $|i$ is used to denote the
covariant derivative associated with the metric $\gamma _{ij}$ induced
on these spatial sections.  The transversality condition for the
vectors and tensor is then written as $S_{|i}^{i}=0$, $F_{|i}^{i}=0$
and $%
h_{j|i}^{i}=0$. The global factor $a^{2}(\eta )$ is introduced for
later convenience, but could be reabsorbed. The metric perturbations
will henceforth be treated linearly.

Four functions describing the metric perturbations are of scalar type,
four more are of vector type and finally there are two which are of
tensor type, according to their transformation properties on the
three-dimensional spatial sections
\cite{bardeen80,halliwell85,stewart90}. These types are preserved by
time evolution provided that the perturbations of the matter sources
around the configuration generating the background Robertson-Walker
geometry are also treated linearly.  Those ten functions do not
characterize in a unique way non-equivalent perturbed geometries since
they may arise not only due to real perturbations of the geometry but
also to changes of the mapping from the background manifold to the
perturbed one.  Hence, a diffeomorphism generated by a vector field
$\vec{\zeta}$, considered to be of the same order as the metric
perturbations, would give an extra contribution
$\mathcal{L}_{\vec{\zeta}}g_{ab}$ to the metric perturbation $h_{ab}$,
where $g_{ab}$ is the background metric.  These local diffeomorphisms
do not preserve in general the scalar, vectorial or tensorial nature
of the metric perturbations.

There are different approaches to overcome the difficulties derived
from this \emph{gauge} freedom.  One approach is to \emph{fix the
  gauge} \cite{lifshitz46} so that further changes on the metric
perturbations resulting from coordinate changes, are not allowed. This
can be achieved by fixing some of the ten functions characterizing the
components of the metric perturbations either directly specifying some
components of $h_{ab}$ or imposing relations between them.  A second
approach, first used by Bardeen \cite{bardeen80}, is based on the
introduction of so-called \emph{gauge-invariant variables}
%(this formalism has been mainly applied to scalar-type perturbations)
which corresponds to using linear combinations of those ten functions
which remain invariant to linear order under diffeomorphisms generated
by any vector field $\vec{\zeta}$. One can always argue that those
gauge-invariant variables coincide with the value taken by the
functions appearing in Eq.~(\ref{8.20}) (or some linear combination of
them) in some particular gauge, as follows from the remark that the
components of any tensorial object referred to a particular and fixed
coordinate system do not change when are reexpressed in terms of some
new coordinates \cite{unruh98}.

{}From now on we will consider spatially flat Robertson-Walker
metrics, \emph{i.e.} $\gamma_{ij}=\delta_{ij}$ in Eq.~(\ref{7.101}),
and concentrate on scalar-type metric perturbations. The motivation
for the latter is that scalar-type metric perturbations are the only
ones which couple to matter sources characterized by scalar functions
when both metric perturbations and matter perturbations (the inflaton
perturbations in our case) are treated linearly. Only two true
\emph{kinematical} degrees of freedom (\emph{i.e.}, before imposing
the Einstein equation) exist for this type of perturbations, in the
sense that from the four arbitrary functions characterizing scalar
metric perturbations, the equivalence classes invariant under local
diffeomorphism transformations are completely characterized by two
arbitrary functions \cite{kodama84,stewart90,mukhanov92,mukhanov05}.
A particular example corresponds to the following two linear
combinations of the four functions $\bar\phi $, $\psi$, $B$ and $E$,
which are invariant under local diffeomorphisms:
\begin{eqnarray*}
\Phi  &=&\bar\phi +\frac{1}{a}\left[ \left( B-E^{\prime }\right) a\right]
^{\prime }\text{,} \\
\Psi  &=&\psi -\frac{a^{\prime }}{a}\left( B-E^{\prime }\right) \text{.}
\end{eqnarray*}
These gauge-invariant variables were first introduced by
Bardeen \cite{bardeen80} with the notation
$\Phi _{A}=\Phi $ and $\Phi _{H}=-\Psi $.

One can also define a gauge-invariant version of the linear
perturbations of the Einstein tensor, $(G_\mathrm{inv}^{(1)})_{a}^{b
}$, which depends only on the gauge-invariant functions $\Phi $ and
$\Psi $, and is invariant under the same kind of local diffeomorphism
which preserve the scalar nature of the metric perturbations
characterized by $\Phi $ and $\Psi$; see
Ref.~\cite{mukhanov92,mukhanov05} for details.  In fact, the
gauge-invariant perturbations $(G_\mathrm{inv}^{(1)}) _{a }^{b }$ of
the Einstein tensor coincide with the actual components of the linear
perturbations of the Einstein tensor in the so-called
\emph{longitudinal gauge}, which corresponds to taking
$E=B=0$. Similarly, one could also define a gauge-invariant version of
the stress tensor linear perturbations and write a version of the
Einstein equation for the metric perturbations with both sides
explicitly invariant; recall that the whole linearized Einstein
equation is itself gauge invariant. We will follow an alternative
procedure which yields equivalent results. The idea is to consider the
components of the Einstein equation in the longitudinal gauge and
notice, as will be explicitly shown below, that all the geometric
dependence can be written entirely in terms of the gauge invariant
variables $\Phi$ and $\Psi$ since the only non-vanishing scalar
contributions to the metric perturbations in the longitudinal gauge,
$\bar\phi$ and $\psi$, coincide with $\Phi$ and $\Psi$.

In the longitudinal gauge the expression of the perturbed metric for
scalar type perturbations on a spatially flat Robertson-Walker
background in terms of the two gauge-invariant functions $\Phi (x)$
and $\Psi (x)$ is
\begin{equation}
ds^{2}=a^{2}(\eta )\left[ -(1+2\Phi (x))d\eta ^{2}+(1-2\Psi (x))\delta
_{ij}dx^{i}dx^{j}\right] \text{,}  \label{8.40}
\end{equation}
and the components of the linear perturbation of the Einstein tensor
are
\begin{eqnarray}
G^{(1)\,0}_{\ \ \ 0} &=&2a^{-2}\left( 3\mathcal{H}(\mathcal{H}\Phi
+\Psi ^{\prime })-\nabla ^{2}\Psi \right) \text{,}  \label{8.41} \\
G^{(1)\,i}_{\ \ \ 0} &=&2a^{-2}\partial _{i}(\mathcal{H}\Phi +\Psi
^{\prime })\text{,}  \label{8.42} \\
G^{(1)\,j}_{\ \ \ i}&=&2a^{-2}\left[ \left( 2\mathcal{H}^{\prime
}+\mathcal{H}^{2}\right) \Phi +\mathcal{H}\Phi ^{\prime }+\Psi ^{\prime
\prime }+2\mathcal{H}\Psi ^{\prime }+\frac{1}{2}\nabla ^{2}D\right] \delta
_{i}^{j} 
-a^{-2}\delta ^{jk}\partial _{k}\partial _{i}D\text{,}  \label{8.43}
\end{eqnarray}
where $\mathcal{H}=a^{\prime }(\eta )/a(\eta )$, $D=\Phi -\Psi $,
$\nabla^2 = \delta^{ij}\partial_i\partial_i$ and, as we mentioned
above, primes denote derivation with respect to the conformal time
$\eta $. The Einstein equation for the linear perturbations of the
metric is
\begin{equation}
G^{(1)\,b}_{\ \ \ a}=\kappa \left(T^{(0)\, b}_{\ \ \ a } - \langle T^{(0)\,b}_{\ \ \ a}
\rangle + T^{(1)\,b}_{\ \ \ a}\right)\text{,}
\label{8.44}
\end{equation}
where the right-hand side corresponds to those terms which are not
included in the background stress tensor $\mathcal{T}_{a}^{b}$ and are
at most linear in the metric perturbations (some terms from the stress
tensor on the background geometry, $T^{(0)\,b}_{\ \ \ a}$, are present
because the scalar field is also perturbed). Note that at the
classical level one should simply substitute the unperturbed stress
tensor ${\cal{T}}^b_a$ for $\langle T^{(0)\,b}_{\ \ \ a}
\rangle$. However, $\langle T^{(0)\,b}_{\ \ \ a} \rangle$ has
non-trivial additional components in stochastic gravity or when
quantizing both the metric perturbations and the scalar field; see the
last paragraph of Appendix~\ref{appA} for additional discussion on the
notation and related points. Furthermore, from now on we will only
consider terms which are linear in either the metric perturbations or
the inflaton field perturbations $\varphi$, which are both assumed to
be of the same order. The contribution to the stress tensor of these
linear perturbations will be denoted by $\delta
\mathcal{T}_{a}^{b}$. Hence, taking all this into account, the
components of the Einstein equation for linear scalar perturbations of
the metric become
\begin{equation}
G^{(1)\,b}_{\ \ \ a}=\kappa \delta \mathcal{T}_{a}^{b}\text{.}
\label{8.47}
\end{equation}
Let us remember that the expression for the stress tensor of the free
massive field $\varphi$, which is minimally coupled to the spacetime
curvature and evolves on the perturbed metric
$\tilde{g}_{ab}=g_{ab}+h_{ab}$, is
\begin{equation}
T_{ab}=\tilde{\nabla} _{a}\varphi \tilde{\nabla} _{b}\varphi -\frac{1}{2}
\tilde{g}_{ab}\left( \tilde{g}^{cd}\tilde{\nabla} _{c}\varphi \tilde{\nabla}
_{d}\varphi +m^{2}\varphi ^{2}\right) \text{.}  \label{8.48}
\end{equation}
The components for the linear perturbations of the stress tensor,
$\delta \mathcal{T}_{a}^{b}$, in the basis associated with the
conformal time and comoving spatial coordinates are then
straightforwardly obtained:
\begin{eqnarray}
\delta \mathcal{T}_{0}^{0}&=&a^{-2}\left( \Phi (\phi ^{\prime })^{2}-\phi'
\varphi ^{\prime }-m^{2}a^{2}\phi \varphi \right) \text{,}  \label{8.49} \\
\delta \mathcal{T}_{0}^{i}&=&a^{-2}\phi ^{\prime }\partial _{i}\varphi
\text{,}
\label{8.50} \\
\delta \mathcal{T}_{i}^{j}&=&a^{-2}\delta _{i}^{j}\left( -\Phi (\phi ^{\prime
})^{2}+\phi ^{\prime }\varphi ^{\prime }-m^{2}a^{2}\phi \varphi \right)
\text{.}  \label{8.51}
\end{eqnarray}
Taking into account the fact that $\delta\mathcal{T}_{i}^{j}$ is
diagonal, one can use Eq.~(\ref{8.47}) [see also Eq. (\ref{8.43})]
with $i\neq j$ to conclude that $\Phi $ and $\Psi $ are equal
\cite{mukhanov92,mukhanov05}, except for a possible homogeneous
component (independent of the spatial coordinates), which should be
included in the background scale factor.  Alternatively, the same
conclusion can also be reached by considering the sum of all the
diagonal elements of the $ij$-components of Eq.~(\ref{8.47}) together
with Eqs.~(\ref{8.50}) and (\ref{8.51}) to substitute $\varphi $ in
terms of $\Phi $ and $\Psi $, which yields $\nabla ^{2}\left(
\Phi -\Psi \right) =0$ and hence $\Psi =\Phi $ provided that they
vanish at infinity.

{}From the Friedmann equations (\ref{8.8}) and (\ref{8.9}) for the
background solution, we have
\begin{equation}
\frac{\kappa }{2} (\phi ^{\prime })^{2}=\mathcal{H}^{2}-\mathcal{H}%
^{\prime }\text{,}  \label{8.52}
\end{equation}
and we can use this equation to reexpress the terms in
Eqs.~(\ref{8.49})-(\ref{8.51}) which are linear in $\Phi
$. Substituting these terms into the perturbed Einstein equations,
given by Eq.~(\ref{8.47}), we finally get
\begin{eqnarray}
\nabla ^{2}\Phi -3\mathcal{H}\Phi ^{\prime }-\left( \mathcal{H}^{\prime }
+2\mathcal{H}^{2}\right) \Phi &=&\frac{\kappa }{2}\left( \phi ^{\prime}
\varphi ^{\prime }+m^{2}a^{2}\phi \varphi \right) \text{,}  \label{8.53} \\
\partial _{i}(\Phi ^{\prime }+\mathcal{H}\Phi ) &=&\frac{\kappa }{2}\phi
^{\prime }\partial _{i}\varphi \text{,}  \label{8.54} \\
\left( \mathcal{H}^{\prime }+2\mathcal{H}^{2}\right) \Phi +3\mathcal{H}\Phi
^{\prime }+\Phi ^{\prime \prime } &=&\frac{\kappa }{2}\left( \phi ^{\prime
}\varphi ^{\prime }-m^{2}a^{2}\phi \varphi \right) \text{.}  \label{8.55}
\end{eqnarray}

Similarly, the Klein-Gordon equation for the inflaton perturbations
can be obtained by linearizing in both the metric perturbations $\Phi
$ and the inflaton perturbations $\varphi $ the exact Klein-Gordon
equation for the whole inflaton field $\phi (\eta )+\varphi (x)$ on
the perturbed geometry, Eq.~(\ref{8.21b}), and making use of the fact
that the homogeneous background solution $\phi (\eta )$ satisfies the
Klein-Gordon equation on the background spacetime, Eq.~(\ref{8.10}).
It can also be obtained from the conservation equation, to linear
order in the metric perturbations, of the linearly perturbed stress
tensor. The result is
\begin{equation}
\varphi ^{\prime \prime }+2\mathcal{H}\varphi ^{\prime }-\nabla ^{2}\varphi
+m^{2}a^{2}\varphi -4\phi ^{\prime }\Phi ^{\prime }+2m^{2}a^{2}\phi \Phi =0%
\text{,}  \label{8.56}
\end{equation}
where we have already taken into account that $\Psi =\Phi $.

\subsection{Einstein-Langevin equation for linearized cosmological
perturbations}

\label{sec3.2}

Using the results of the previous subsection, the Einstein-Langevin
equation (\ref{7.7}) can be particularized to the case of scalar-type
metric perturbations around a Robertson-Walker background geometry
with the following result:
\begin{eqnarray}
\frac{\kappa }{2}a^{2}\left( \langle \delta \hat{\mathcal{T}}_{0}^{0}
\rangle _{\Phi }+\xi _{0}^{0}\right) &=&3\mathcal{H}(\mathcal{H}\Phi
+\Psi^{\prime })-\nabla ^{2}\Psi \text{,}  \label{8.70} \\
\frac{\kappa }{2}a^{2}\left(\langle \delta \hat{\mathcal{T}}_{0}^{i}
\rangle _{\Phi }+\xi _{0}^{i}\right) &=&\partial _{i}(\Psi ^{\prime }+%
\mathcal{H}\Phi )\text{,}  \label{8.71} \\
\frac{\kappa }{2}a^{2}\left(\langle \delta \hat{\mathcal{T}}_{i}^{j}
\rangle _{\Phi }+\xi _{i}^{j}\right) &=&\left[ \left( 2\mathcal{H}^{\prime }
+%
\mathcal{H}^{2}\right) \Phi +\mathcal{H}\Phi ^{\prime }+\Psi ^{\prime \prime
}+2\mathcal{H}\Psi ^{\prime }+\frac{1}{2}\nabla ^{2}D\right] \delta _{i}^{j}
-\frac{1}{2}\delta ^{jk}\partial _{k}\partial _{i}D\text{.}  \label{8.71b}
\end{eqnarray}
where we have used Eqs.~(\ref{8.41})-(\ref{8.43}) for the linearized
Einstein tensor and we have considered the stress tensor operator
$\delta \hat{\mathcal{T}}_{a}^{b}$, which results from keeping terms
linear in either the inflaton perturbations or the metric
perturbations. The notation $\langle \delta
\hat{\mathcal{T}}_{a}^{b}\rangle _{\Phi }$ is equivalent to $\langle
\delta \hat{\mathcal{T}}_{a}^{b}[g+h]\rangle$ for the particular case
of scalar metric perturbations that we are considering. Note, in
addition, that all the contributions to $\langle \delta
\hat{\mathcal{T}}_{a}^{b} \rangle _{\Phi }$ are, explicitly or
implicitly, proportional to the metric perturbations since otherwise
they would be proportional to $\langle \hat{\varphi}[g]\rangle $,
which vanishes.  In fact, this turns out to be important so that
$\langle \delta \hat{\mathcal{T}}_{a}^{b}[g+h]\rangle$ coincides with
$\langle \hat{T}^{(1)\,b}_{\ \ \ a}[g+h]\rangle$, which is the object
that appears in the Einstein-Langevin equation, when one keeps to
linear order in the inflaton perturbations, \emph{i.e.}, when only the
first two terms on the right-hand side of Eq.~(\ref{8.21}) are
considered.

The three equations corresponding to the spatial components with equal
indices of the Einstein-Langevin equation are equivalent due to the
symmetries of the Robertson-Walker metric and those of the Gaussian
state of the inflaton perturbations being considered, which was chosen
to be compatible with those symmetries. On the other hand, the
equation for the spatial components with different indices can be used
in a similar way to that of the previous subsection in order to show
that the gauge invariant functions for the scalar metric perturbations
$\Phi (x)$ and $\Psi (x)$ coincide. In this case it is also necessary
that the spatial components of the stochastic source $\xi _{ij}(x)$
with indices $i\neq \!j\;$ vanish identically.  Indeed, since $\xi
_{ab}(x)$ is a Gaussian stochastic process with zero mean, $\xi
_{ij}(x)$ will vanish provided that $\left\langle \xi _{ij}(x)\xi
  _{cd}(y)\right\rangle _{\xi }=0$, which can be argued as
follows. The correlation function for $\xi_{ab}$ is defined by the
noise kernel and, as we are keeping to linear order in the inflaton
perturbations, only the first contribution in Eq.~(\ref{8.22}),
$\left\langle \left\{ \hat{t}_{ij}[g],\hat{t}_{cd}[g]
  \right\}\right\rangle _{\phi \varphi }=\left\langle \left\{ \delta
    \hat{t} _{ij}[g],\delta \hat{t}_{cd}[g]\right\}\right\rangle$,
where $\delta\hat{t} _{ab}\equiv\delta \hat{\mathcal{T}}_{ab}-\langle
\delta \hat{\mathcal{T}}_{ab}\rangle$, should be considered. Finally,
$\left\langle\left\{\delta\hat{t}_{ij}[g],\delta\hat{t}_{cd}[g]\right\}
\right\rangle$ vanishes for $i \neq \!j$ since $\delta
\hat{\mathcal{T}}_{ij}=0$ in that case, as follows from
Eq.~(\ref{8.51}) with the inflaton perturbation promoted to a quantum
operator. Hence, from now on we will take $\Psi
=\Phi$. Eq.~(\ref{8.71b}) is then trivially satisfied for $i\neq
\!j\;$ and for $i=\!j\,$ it reduces to (no summation should be
understood over the repeated index $i$)
\begin{equation}
\frac{\kappa }{2}a^{2}\left( \langle \delta \hat{\mathcal{T}}_{i}^{i}
\rangle _{\Phi }+\xi _{i}^{i}\right) =\left( 2\mathcal{H}^{\prime }
+\mathcal{H}^{2}\right) \Phi +3\mathcal{H}\Phi ^{\prime }+\Phi ''
\text{.}  \label{8.71c}
\end{equation}
It is clear that Eqs.~(\ref{8.70})-(\ref{8.71b}) are redundant since
we have three equations but only two variables to be determined: the
function $\Phi $ characterizing the metric perturbations of scalar
type and the expectation value of the quantum operator for the
inflaton perturbations on the spacetime with the perturbed metric,
$\left\langle \hat{\varphi}[g+h]\right\rangle $, which will also be
denoted in this case by $\left\langle \hat{\varphi} \right\rangle
_{\Phi }$. However, despite the apparently excessive number of
equations, the system is integrable and solutions can be found. This
fact is guaranteed by the Bianchi identity provided that the source of
the Einstein-Langevin equation is conserved. This is indeed the case:
the averaged and stochastic sources are separately conserved. On the
one hand, the conservation of $\langle \delta
\hat{\mathcal{T}}_{ab}\rangle _{\Phi }$ is equivalent to the
Klein-Gordon equation for the expectation value $\left\langle
\hat{\varphi}\right\rangle _{\Phi }$, which is completely analogous
to Eq.~(\ref{8.56}):
\begin{equation}
\left\langle \hat{\varphi}\right\rangle _{\Phi }^{\prime \prime }
+2\mathcal{H}\left\langle \hat{\varphi}\right\rangle _{\Phi }^{\prime }-\nabla
^{2}\left\langle \hat{\varphi}\right\rangle _{\Phi }+m^{2}a^{2}\left\langle
\hat{\varphi}\right\rangle _{\Phi }-4\phi ^{\prime }\Phi
^{\prime}+2m^{2}a^{2}\phi \Phi =0 \text{.}  \label{8.71e}
\end{equation}
On the other hand, the conservation of the stochastic source is a
consequence of the conservation of the noise kernel, which in turn
relies on the fact that the quantum operator for the inflaton
perturbations $\hat{\varphi}[g]$ satisfies the Klein-Gordon equation
on the background spacetime, $\left(\nabla_a\nabla^a-m^{2}\right)
\hat{\varphi}(x)=0$.

Taking all these considerations into account, the Klein-Gordon
equation (\ref {8.71e}) can be used to obtain the expectation value
$\left\langle \hat{\varphi}\right\rangle _{\Phi }$ in terms of $\Phi
$. We can then easily write the expectation value of the stress tensor
linear perturbations $\langle \delta
\hat{\mathcal{T}}_{a}^{b}\rangle_{\Phi}$ in terms of $\Phi $ and use
any of the constraint equations, Eq.~(\ref{8.70}) or (\ref{8.71}) to
express $\Phi $ entirely in terms of the stochastic source $\xi
_{ab}$; to be specific, in this subsection we will consider
Eq.~(\ref{8.71}).  The spatial derivatives can be easily handled by
working in Fourier space. Hence, in the rest of this section we will
work with Fourier transformed expressions in the spatial coordinates.
A subindex $k$ will denote the three-dimensional comoving momentum
vector $\vec{k}$ that labels each Fourier mode in flat space,
\emph{i.e.},
\begin{equation}
\Phi _{k}(\eta )=\int d^{3}xe^{-i\vec{k}\cdot \vec{x}}\Phi \left( \eta
,\vec{%
x}\right) \text{.}  \label{8.71f}
\end{equation}
Thus, the Fourier-transformed version of Eq.~(\ref{8.71}) is
\begin{equation}
ik_{i}(\Phi _{k}^{\prime }+\mathcal{H}\Phi _{k})=\frac{\kappa }{2}%
a^{2}\left( \langle ( \delta \hat{\mathcal{T}}_{0}^{i})
_{k}\rangle _{\Phi }+( \xi _{0}^{i}) _{k}\right) \text{,}
\label{8.71g}
\end{equation}
where $k_{i}$ is the comoving momentum component associated with the
comoving coordinate $x^{i}$.

Since $\langle \delta \hat{\mathcal{T}}_{a}^{b}\rangle _{\Phi }$ is
linear in $\Phi $, Eq.~(\ref{8.71g}) is a first order linear
integro-differential equation with an inhomogeneous term corresponding
to the $0i$ component of the stochastic source $\xi_{a}^{b}$. Therefore,
one can always write the solution to
Eq.~(\ref{8.71g}) for $\eta \geq \eta_0$ as
\begin{equation}
\Phi_{k}(\eta)=\Phi^\mathrm{(h)}_k (\eta)+\Phi^\mathrm{(i)}_k (\eta)
=\Phi^\mathrm{(h)}_k (\eta) + \frac{\kappa}{2}\int^\eta _{\eta_0}d\eta'
G_\mathrm{ret}^{(k)}(\eta ,\eta ^{\prime })a^{2}\left( \eta ^{\prime }\right)
\left( \xi _{0}^{i}\right) _{k}(\eta ^{\prime })\text{,}  \label{8.72}
\end{equation}
where $\Phi^\mathrm{(h)}_k$ is the solution to the homogeneous version
of Eq.~(\ref{8.71g}) with some given initial conditions at an initial
time $\eta_0$, $\Phi^\mathrm{(i)}_k$ is a solution of the inhomogeneous
equation which vanishes at $\eta_0$ and $G_\mathrm{ret}^{(k)}$ is the
retarded propagator associated with Eq.~(\ref{8.71g}). The correlation
function for the scalar metric perturbation regarded as a solution of
the stochastic differential equation (\ref{8.71g}) corresponds to
\begin{equation}
\langle \Phi _{k}(\eta _{1})\Phi _{k'}(\eta _{2})\rangle_{\xi }
= \langle\Phi^\mathrm{(h)}_{k}(\eta_{1})\Phi^\mathrm{(h)}_{k'}(\eta_{2})
\rangle_{\xi}
+ \langle\Phi^\mathrm{(h)}_{k}(\eta_{1})\Phi^\mathrm{(i)}_{k'}(\eta_{2})
\rangle_{\xi}
+\langle\Phi^\mathrm{(i)}_{k}(\eta_{1})\Phi^\mathrm{(h)}_{k'}(\eta_{2})
\rangle_{\xi}
+ \langle\Phi^\mathrm{(i)}_{k}(\eta_{1})\Phi^\mathrm{(i)}_{k'}(\eta_{2})
\rangle_{\xi} , \label{8.72b}
\end{equation}
where $\left\langle \dots \right\rangle _{\xi }$ denotes the average
over all possible realizations of the stochastic source, as previously
defined.  {}From now on we will concentrate solely on the last term,
which comes entirely from the solutions of the inhomogeneous equation;
see Appendix~\ref{appC} for a discussion on the role of the initial
conditions and the contributions of the homogeneous solution to the
correlation function. The correlation function has then the following
form:
\begin{eqnarray}
\langle \Phi^\mathrm{(i)}_{k}(\eta_{1})\Phi^\mathrm{(i)}_{k'}(\eta_{2})
\rangle_{\xi} &=&\left( \frac{\kappa }{2}\right) ^{2}\int_{\eta_0}^{\eta_1} d\eta
_{1}^{\prime }\int_{\eta_0}^{\eta_2} d\eta _{2}^{\prime } \, a^{2}\left( \eta _{1}^{\prime }\right)
a^{2}\left( \eta _{2}^{\prime }\right) G_\mathrm{ret}^{(k)}(\eta _{1},\eta
_{1}^{\prime })  \nonumber
\\
&&\left\langle \left( \xi _{0}^{i}\right) _{k}(\eta _{1}^{\prime })\left(
\xi _{0}^{i}\right) _{k^{^{\prime }}}(\eta _{2}^{\prime })\right\rangle
_{\xi }G_\mathrm{ret}^{(k^{\prime })}(\eta _{2},\eta _{2}^{\prime })  \nonumber \\
&=&\left( \frac{\kappa }{2}\right) ^{2}\left( G_\mathrm{ret}^{(k)}\cdot \left(
N_{0i0i}\right) _{kk^{^{\prime }}}\cdot ( G_\mathrm{ret}^{(k^{\prime
})}) ^T \right) (\eta _{1},\eta _{2})\text{,}  \label{8.73}
\end{eqnarray}
where we used the notation $A^{T} (\eta ,\eta')=A(\eta',\eta )$ and $A
\cdot B = \int_{\eta_0}^\infty d\eta \, A(\eta)B(\eta)$, and the
factors $a^{2}(\eta'_1)$ and $a^{2}(\eta'_2)$ were simplified when
lowering the spatial indices with the background metric in the last
equality.  Since we are linearizing in the inflaton perturbations,
only the first term on the right-hand side of Eq.~(\ref{8.22}) should
be considered. The expression for the Fourier-transformed version of
the noise kernel then becomes
\begin{equation}
\left( N_{abcd}\right) _{kk^{^{\prime }}}(\eta,\eta')=\frac{1}{2}
\left\langle \left\{ \left(\delta\hat{t}_{ab}\right)_{k}(\eta),\left( \delta
\hat{t}_{cd}\right) _{k'}(\eta')\right\} \right\rangle _{\Phi=0}\text{,}
\label{8.74}
\end{equation}
where $\delta\hat{t}_{ab}= \delta\hat{\mathcal{T}}_{ab} -
\langle\delta\hat{\mathcal{T}}_{ab} \rangle$, as defined earlier, and
$\left\langle \ldots \right\rangle _{\Phi =0}$ is the expectation
value for the product of quantum operators $\delta\hat{t}_{ab}$ with
the field $\hat{\varphi}$ evolving on the background metric. We
finally obtain the following expression relating the correlation
function for the metric perturbations and the fluctuations of the
stress tensor operator:
\begin{equation}
\langle \Phi^\mathrm{(i)}_{k}(\eta _{1})\Phi^\mathrm{(i)}_{k'}(\eta_{2})\rangle
_{\xi }=\left( \frac{\kappa }{2}\right) ^{2}\frac{1}{2}
\left( G_\mathrm{ret}^{(k)}\cdot \left\langle \left\{ \left( \delta\hat{t}_{0i}
\right)_{k},\left(\delta\hat{t}_{0i}\right) _{k'}\right\} \right\rangle
_{\Phi =0}\cdot ( G_\mathrm{ret}^{(k')}) ^{T}\right) (\eta_{1},\eta _{2})
\,\text{.}  \label{8.75}
\end{equation}
A detailed example of this kind of computation is given in the next
section, where the correlation function of scalar-type metric
perturbations will be computed for the particular case in which the
background solutions for $\phi(\eta)$ and $a(\eta)$ correspond to a
period of slow-roll inflation.

We end this subsection by working out the explicit expression for the
$0i$ component of the expectation value
$\langle\delta\hat{\mathcal{T}}_{a}^{b}\rangle _{\Phi}$. From
Eq.~(\ref{8.50}) we get
\begin{equation}
\langle ( \delta \hat{\mathcal{T}}_{0}^{i}) _{k}(\eta )
\rangle_{\Phi }=ik_{i}a^{-2}\left( \eta \right) \phi ^{\prime }(\eta )
\left\langle
\hat{\varphi}_{k}(\eta )\right\rangle _{\Phi }\text{,}  \label{8.76}
\end{equation}
and everything reduces to compute the expectation value $\left\langle
\hat{\varphi}_{k}(\eta )\right\rangle _{\Phi }$. One way of
obtaining it is by regarding $\Phi _{k}$ as an external source of the
Fourier-transformed version of the linearized Klein-Gordon equation
(\ref{8.71e}) and solving the corresponding inhomogeneous equation
perturbatively so that $\langle \hat{\varphi}_{k}[g+h] \rangle =
\langle \hat{\varphi}_{k}^{(0)}[g] \rangle + \langle
\hat{\varphi}_{k}^{(1)}[g+h] \rangle + O (h^2)$.
%up to linear order in $h_{ab}$.
The expectation value $\langle \hat{\varphi}_{k}^{(0)}[g]\rangle$
vanishes, and $\langle \hat{\varphi}_{k}^{(1)}[g+h] \rangle$ is the
solution of the inhomogeneous equation with vanishing initial
conditions, which is proportional to the metric perturbation $\Phi
_{k}$ and can be written as
%Here
%$\hat{\varphi}_{k}^{(0)}[g]$ is a solution of the Klein-Gordon equation on
%the background metric (the homogeneous part of Eq.~(\ref{8.71e})) and
%includes the entire specification of the initial conditions at $\eta _{0}$,
%which will provide the information about the initial state of the inflaton
%perturbations when taking the expectation value with respect to their
%corresponding time-independent quantum state in the Heisenberg picture. On
%the other hand, $\hat{\varphi}_{k}^{(1)}[g+h]$ is the solution of the
%inhomogeneous equation with vanishing initial conditions, which is
%proportional to the metric perturbation $\Phi _{k}$. Being evaluated on the
%unperturbed geometry, the expectation value $\langle \hat{\varphi}_{k}
%^{(0)}[g]\rangle$ vanishes and $\hat{\varphi}_k^{(1)}$
%can be written as
\begin{equation}
\langle \hat{\varphi}_{k}^{(1)}(\eta ) \rangle
= \int_{\eta_0}^{\eta} d\eta ^{\prime
}\bar{G}_\mathrm{ret}^{(k)} (\eta ,\eta ^{\prime })\biggl( 4\phi^{\prime}(\eta')
\Phi _{k}^{\prime }(\eta ^{\prime })-2m^{2}a^{2}(\eta ^{\prime })\phi
(\eta ^{\prime })\Phi _{k}(\eta ^{\prime })\biggr) \text{,}  \label{8.77}
\end{equation}
where $\bar{G}_\mathrm{ret}^{(k)}$ is the Fourier-transformed version
of the retarded propagator associated with the Klein-Gordon equation
(\ref{8.71e}) with vanishing initial conditions at $\eta _{0}$. One
can show that the retarded propagator $\bar{G}_\mathrm{ret}^{(k)}$ for
the Klein-Gordon equation (\ref{8.71e}) with Fourier-transformed
spatial coordinates is given by
\begin{equation}
\bar{G}_\mathrm{ret}^{(k)}(\eta ,\eta ^{\prime })=ia^{2}(\eta')
\left\langle \left[ \hat{\varphi}_{k}(\eta ),\hat{\varphi}_{-k}(\eta
^{\prime })\right] \right\rangle \theta (\eta -\eta ^{\prime})
\text{,}  \label{8.78}
\end{equation}
%where the factor $\theta (\eta ^{\prime }-\eta _{0})$ was included to ensure
%vanishing initial conditions at $\eta _{0}$. Altogether, we get the
%following result for the expectation value $\langle ( \delta
%\hat{\mathcal{T}}_{0}^{i}) _{k}\rangle _{\Phi }$:
Substituting Eq.~(\ref{8.78}) into Eq.~(\ref{8.77}) and the result
into Eq.~(\ref{8.76}) one gets the following result for the
expectation value $\langle ( \delta \hat{\mathcal{T}}_{0}^{i})
_{k}\rangle _{\Phi }$:
\begin{equation}
\langle ( \delta \hat{\mathcal{T}}_{0}^{i}) _{k}(\eta )
\rangle
_{\Phi } = ik_{i}a^{-2} (\eta) \phi ^{\prime }(\eta )\int_{\eta
_{0}}^{\eta }d\eta ^{\prime }i\left\langle \left[ \hat{\varphi}_{k}(\eta ),
\hat{\varphi}_{-k}(\eta ^{\prime })\right] \right\rangle a^{2}(\eta')
\biggl( 4\phi' (\eta ^{\prime })\Phi _{k}^{\prime }(\eta ^{\prime })
%\nonumber \\
%&&
-2m^{2}a^{2}(\eta ^{\prime })\phi (\eta ^{\prime })\Phi _{k}(\eta
^{\prime })\biggr) \text{.}  \label{8.79}
\end{equation}
Note that this expression for the expectation value of the stress
tensor operator requires no renormalization because we linearized with
respect to the scalar field perturbations; see Appendix~\ref{appB} for
further comments on this point. Furthermore, in Appendix~\ref{appB} we
also show that the expectation value obtained above is in agreement
with the general expression for the expectation value of the stress
tensor which follows from the approach to the Einstein-Langevin
formalism based on functional methods.

\subsection{Equivalence with the usual quantization methods}

\label{sec3.3}

In this subsection we will show that the result for the correlation
function of the metric perturbations obtained in the previous
subsection using the Einstein-Langevin equation and linearizing in the
inflaton perturbations coincides with the result which follows from
the usual quantization procedures in linear cosmological perturbation
theory; see for instance Refs.~\cite{mukhanov92,kodama84,mukhanov05}.

Let us promote the scalar-type metric perturbations $\Phi$ and the
inflaton perturbations $\varphi $ to quantum
operators. Eqs.~(\ref{8.53})-(\ref{8.55}) then become equations for
the operators in the Heisenberg picture. In particular, we will
concentrate on the temporal components of the Einstein equation
\begin{eqnarray}
3\mathcal{H}\hat{\Phi}^{\prime } + 3\mathcal{H}^{2} \hat{\Phi}
-\nabla ^{2}\hat{\Phi} &=&\frac{\kappa }{2}a^{2}
\delta \hat{\mathcal{T}}_{0}^{0}| _{\hat{\Phi}}\text{,}  \label{8.90}
\\
\partial _{i}(\hat{\Phi}^{\prime }+\mathcal{H}\hat{\Phi}) &=&\frac{\kappa }
{2%
}a^{2} \delta \hat{\mathcal{T}}_{0}^{i}| _{\hat{\Phi}}\text{,}
\label{8.91}
\end{eqnarray}
where the quantum operator for the inflaton perturbations
$\hat{\varphi}[g+h]$, on which $\delta \hat{\mathcal{T}}_{ab}|
_{\hat{\Phi}}$ depends, satisfies the linearized Klein-Gordon equation
\begin{equation}
\hat{\varphi}^{\prime \prime }+2\mathcal{H}\hat{\varphi}^{\prime }-\nabla
^{2}\hat{\varphi}+m^{2}a^{2}\hat{\varphi}-4\phi ^{\prime }\hat{\Phi}^{\prime
}+2m^{2}a^{2}\phi \hat{\Phi}=0\text{.}  \label{8.92}
\end{equation}
The situation is completely analogous to that of the previous
subsection except for the fact that the metric perturbation
$\hat{\Phi}(x)$ is now a genuine quantum operator instead of a
stochastic $c$-number. Thus, taking the Fourier transform for the
spatial coordinates and proceeding in a similar fashion to the
previous subsection, the Klein-Gordon equation~(\ref{8.92}) can be
solved with the following result:
\begin{equation}
\hat{\varphi}_{k}(\eta )=\hat{\varphi}_{k}^{(0)}(\eta )+\int d\eta ^{\prime}
\bar{G}_\mathrm{ret}^{(k)}(\eta ,\eta ^{\prime })\left( 4\phi ^{\prime }(\eta
^{\prime })\hat{\Phi}_{k}^{\prime }(\eta ^{\prime })-2m^{2}a^{2}(\eta
^{\prime })\phi (\eta ^{\prime })\hat{\Phi}_{k}(\eta ^{\prime })\right)
\text{,}  \label{8.93}
\end{equation}
where $\hat{\varphi}^{(0)}_k(\eta)$ is a solution of the Klein-Gordon
equation on the background spacetime which contains the entire
specification of the initial conditions at time $\eta _{0}$ and
$\bar{G}_\mathrm{ret}^{(k)}$ is the retarded propagator with vanishing
initial conditions associated with the Fourier-transformed version of
Eq.~(\ref{8.92}), which coincides with that already obtained for
Eq.~(\ref{8.71e}).

Taking into account Eq.~(\ref{8.93}), one could use any of the
constraint equations (\ref{8.90}) or (\ref{8.91}) to express the
quantum operator for the metric perturbation $\hat{\Phi}$ entirely in
terms of the operator for the inflaton perturbations
$\hat{\varphi}^{(0)}[g]$ (in addition to the scalar functions $\phi
(\eta)$ and $a(\eta)$ characterizing the background solution), which
satisfies the Klein-Gordon equation on the unperturbed
geometry. However, before proceeding further it is convenient to
discuss some useful expressions relating $\delta
\hat{\mathcal{T}}_{ab}$ and its expectation values on both the
background spacetime and the perturbed geometry. The stress tensor
operator $\delta \hat{\mathcal{T}}_{ab}|_{\hat{\Phi}}$, which is
linear in both $\hat{\varphi}[g+h]$ and $\hat{\Phi}$, can actually be
written as a linear combination of terms proportional to
$\hat{\varphi}^{(0)}[g]$, the inflaton perturbations on the background
metric, and terms proportional to $\hat{\Phi}$. The latter correspond
to terms coming either from the explicit dependence of the stress
tensor on the metric, which give a local contribution, or from the
dependence of $\hat{\varphi} [g+h]$ on the metric perturbations
according to Eq.~(\ref{8.93}).  In fact, since Eq.~(\ref{8.93}) is
identical to Eq.~(\ref{8.77}) when substituting the stochastic
function $\Phi_k(\eta)$ by the operator $\hat{\Phi}_k(\eta)$, it is
clear that all the terms proportional to $\hat{\Phi}$ in $\delta
\hat{\mathcal{T}}_{ab}|_{\hat{\Phi}}$ are identical to the terms
proportional to $\Phi$ in the operator $\delta
\hat{\mathcal{T}}_{ab}[g+h]$ considered in the previous
subsection. Furthermore, since
$\langle\hat{\varphi}^{(0)}[g]\rangle=0$, those terms were identical
to $\langle \delta\hat{\mathcal{T}}_{ab}\rangle_{\hat{\Phi}}$, where
$\langle \delta\hat{\mathcal{T}}_{ab}\rangle_{\hat{\Phi}}$ should be
understood as the result of replacing $\Phi$ with $\hat{\Phi}$ in the
expectation value $\langle \delta\hat{\mathcal{T}}_{ab}\rangle_{\Phi}$
of the previous subsection. Hence, we have
\begin{equation}
 \delta \hat{\mathcal{T}}_{ab}| _{\hat{\Phi}}-\langle \delta
\hat{\mathcal{T}}_{ab}\rangle _{\hat{\Phi}}=\delta
\hat{\mathcal{T}}_{ab}| _{\hat{\Phi}=0}= \delta \hat{\mathcal{T}}
_{ab}| _{\hat{\Phi}=0}-\langle \delta \hat{\mathcal{T}}_{ab}
\rangle _{\hat{\Phi}=0} \text{,}  \label{8.94}
\end{equation}
where we used in the last equality the fact that $\langle \delta
\hat{\mathcal{T}}_{ab}\rangle _{\hat{\Phi}=0}$ actually vanishes.
Eq.~(\ref{8.94}) can be written as
\begin{equation}
 \delta \hat{\mathcal{T}}_{ab}| _{\hat{\Phi}}=\delta\hat{t}_{ab}
+\langle \delta \hat{\mathcal{T}}_{ab}\rangle _{\hat{\Phi}}
\text{,} \label{8.95}
\end{equation}
where $\delta\hat{t}_{ab}=
\delta\hat{\mathcal{T}}_{ab}|_{\hat{\Phi}=0} - \langle \delta
\hat{\mathcal{T}}_{ab}\rangle_{\hat{\Phi}=0}$.  It should be remarked
that taking the expectation value with respect to some quantum state
of the field $\varphi$ should be considered with caution here since,
due to the constraint equations (\ref{8.90}) and (\ref{8.91}), the
operators $\hat{\varphi}$ and $\hat{\Phi}$ are not independent. Thus,
strictly speaking, $\langle
\delta\hat{\mathcal{T}}_{ab}\rangle_{\hat{\Phi}}$ should be regarded
in this context merely as a notation for those terms of $\delta
\hat{\mathcal{T}}_{ab}| _{\hat{\Phi}}$ which are proportional to
$\hat{\Phi}$, in contrast to those proportional to
$\hat{\varphi}^{(0)}[g]$.

Substituting Eq.~(\ref{8.95}) into any of the constraint equations
(\ref {8.90}) or (\ref{8.91}), one could easily obtain the metric
perturbation $\hat{\Phi}$ in terms of the operator
$\delta\hat{t}_{ab}$ constructed with operators for the inflaton
perturbations evolving on the background metric.  In order to compare
in detail with the result of the previous subsection, where the
Einstein-Langevin equation was used, we will explicitly consider the
case in which the constraint equation (\ref{8.91}) is used. Having
substituted $\delta \hat{\mathcal{T}}_{ab}| _{\hat{\Phi}}$ by $\delta
\hat{t}_{ab}+\langle \delta\hat{\mathcal{T}}_{ab}\rangle_{\hat{\Phi}}$
into the Fourier-transformed version of Eq.~(\ref{8.91}), we have
\begin{equation}
ik_{i}(\hat{\Phi}_{k}^{\prime }+\mathcal{H}\hat{\Phi}_{k})-\frac{\kappa }{2}%
a^{2}\langle ( \delta \hat{\mathcal{T}}_{0}^{i}) _{k}
\rangle _{%
\hat{\Phi}}=\frac{\kappa }{2}a^{2}\left(\delta \hat{t}_{0}^{i}\right) _{k}\text{,}
\label{8.96}
\end{equation}
Taking into account that $\langle\delta \hat{\mathcal{T}}_{0}^{i}
\rangle _{\hat{\Phi}}$ is linear in $\hat{\Phi}$, one can obtain the
following expression for $\hat{\Phi}$ in terms of
$\delta\hat{t}_{0}^{i}$ from Eq.~(\ref{8.96}):
\begin{equation}
\hat{\Phi}_k(\eta )=\hat{\Phi}^\mathrm{(h)}_k(\eta)+\hat{\Phi}^\mathrm{(i)}_k(\eta),
\end{equation}
where $\hat{\Phi}^\mathrm{(h)}_k$ is a solution of the homogeneous
version of Eq.~(\ref{8.96}) and
\begin{equation}
\hat{\Phi}^\mathrm{(i)}_k(\eta )=\frac{\kappa }{2}\int_{\eta _{0}}^{\eta }d\eta
^{\prime }a^{2}\left( \eta ^{\prime }\right) G_\mathrm{ret}^{(k)}(\eta ,\eta
^{\prime })\left(\delta \hat{t}_{0}^{i}\right) _{k}(\eta')=\frac{\kappa
}{2}\left( G_\mathrm{ret}^{(k)}\cdot \left( \delta\hat{t}_{0i}\right) _{k}\right)
(\eta ) \text{,}  \label{8.97}
\end{equation}
where $G_\mathrm{ret}^{(k)}$ is the retarded Green function
%which vanishes for conformal times previous to $\eta_0$
and the factor $a^2(\eta')$ canceled out when lowering the spatial
index $i$ with the background metric. In fact, since the homogeneous
part of Eq.~(\ref{8.96}) has exactly the same form as that of
Eq.~(\ref{8.71g}), the retarded Green function $G_\mathrm{ret}^{(k)} $
coincides with that of the previous subsection.  An analogous remark
holds for the solution of the homogeneous equation,
$\hat{\Phi}^\mathrm{(h)}_k$.

Using Eq.~(\ref{8.97}) and concentrating on the inhomogeneous
contribution (a discussion of the homogeneous solution and its
relationship to the initial conditions is given in
Appendix~\ref{appC}), the symmetrized two-point quantum correlation
function for the metric perturbation operator $\hat{\Phi}$ can be
written as
\begin{equation}
\frac{1}{2}\langle \{ \hat{\Phi}^\mathrm{(i)}_k(\eta _{1}),\hat{\Phi}
^\mathrm{(i)}_{k'}(\eta _{2})\} \rangle =\left( \frac{\kappa}{2}
\right)^{2}\frac{1}{2}\left( G_\mathrm{ret}^{(k)}\cdot \left\langle \left\{ \left(
\delta\hat{t}_{0i}\right) _{k},\left( \delta\hat{t}_{0i} \right)_{k'} \right\}
\right\rangle _{\Phi =0}\cdot (G_\mathrm{ret}^{(k')}) ^{T}\right) \left(
\eta _{1},\eta_{2}\right) \text{.}  \label{8.98}
\end{equation}
Thus, we can see that the result for the symmetrized quantum
correlation function of the metric perturbations coincides with the
stochastic correlation function (\ref{8.75}) obtained in the previous
subsection using the Einstein-Langevin equation.

We end this section by making a few remarks concerning the issue of
the normalization of cosmological perturbations. In principle, one
could differentiate Eq.~(\ref{8.91}) with respect to the conformal
time (the spatial derivatives can be easily eliminated by working in
Fourier space) and combine it with Eq.~(\ref{8.90}) to obtain a linear
second order differential equation for the metric perturbation
operator $\hat{\Phi}$. However, when trying to quantize a theory
beginning with the equations of motion instead of an action, one faces
a normalization ambiguity which stems from the fact that, although any
pair of actions that differ in a constant factor yield the same
equations of motion (either classical or for quantum operators in the
Heisenberg picture), their corresponding quantum theories are not
completely equivalent. In particular, the quantum correlation
functions for a given state (e.g. the fundamental state) do not
coincide. For a linear theory they actually differ by some power of a
constant factor which is precisely the square root of the
proportionality constant between the two actions. This is the reason
why in Ref.~\cite{mukhanov92}, when quantizing the theory for linear
perturbations, the final action was obtained from the original
linearized action for a scalar field evolving on a metric perturbed
around a given background geometry together with the linearized
Einstein-Hilbert action for the perturbations of that metric. That was
done by using the constraint equations to reduce the whole action to
that for the only true dynamical degrees of freedom.  It was precisely
in order to avoid the normalization ambiguity explained above that
such a procedure, which turns out to be rather cumbersome, was used
instead of working directly with the equations of motion and finding
at the end an action which corresponds to the equation of motion for
the true dynamical degree of freedom.

On the other hand, the method employed in this section is not affected
by such a normalization ambiguity because, as can be seen from
Eqs.~(\ref{8.91}) and (\ref{8.50}), the constraint equation relates the
operator for the metric perturbations to the operator for the inflaton
perturbations on the background metric. The normalization of the
latter operator is already determined by the usual procedure of
quantization starting from the action of a scalar field on a fixed
spacetime geometry. Hence, the key point was to separate the inflaton
perturbation operator $\hat{\varphi}$ satisfying the Klein-Gordon
equation~(\ref{8.92}) into a contribution $\hat{\varphi}^{(0)} [g]$
which can be regarded as the inflaton perturbation evolving on the
fixed background spacetime plus a contribution proportional to the
metric perturbation operator $\hat{\Phi}$, and use then the constraint
equation to express $\hat{\Phi}$ entirely in terms of
$\hat{\varphi}^{(0)}[g]$.  In contrast with the approach of
Ref.~\cite{mukhanov92}, this procedure does not give an explicit
expression for the reduced action or even the equation of motion for
an isolated true dynamical degree of freedom, but it is rather useful
(and sufficient) in order to compare with the results obtained in the
previous subsection by means of the Einstein-Langevin equation.

\section{Particular example: computation of the power spectrum
for large scales in a simple inflationary model}

\label{sec4}

In this section we will apply the method developed in
Sec.~\ref{sec3.2} to studying the particular example of metric
fluctuations induced by the quantum fluctuations of the inflaton field
in the context of a simple model of chaotic inflation corresponding to
a free minimally-coupled massive scalar field. In order to carry out
explicit computations, we will assume that the Robertson-Walker
background geometry is close to the de Sitter geometry. For models
with exponential inflation, as the one being considered here, this
approximation is reasonable during the inflationary period, in which
the so-called \emph{slow-roll} parameters controlling the deviations
from de Sitter geometry are small, but not for later times. One can,
nevertheless, obtain useful results from a cosmological point of view
because those scales which are of cosmological interest at present
correspond to scales which left the horizon during the inflationary
period.  This can be understood as follows. On the one hand, the
evolution of gravitational perturbations outside the horizon is fairly
simple, as can be understood from causality arguments
\cite{bardeen83,mukhanov92}, and rather independent of the particular
dynamics of the matter sources. On the other hand, the evolution when
the scale reenters the horizon later on during the radiation and
matter dominated eras has been widely studied using the Newtonian
approximation \cite {kolb90,padmanabhan93}.

We stress that the results obtained in this section, which are based
on the use of the Einstein-Langevin equation, are not new. They are
basically in agreement with most of the literature based on the
simultaneous quantization of gravitational perturbations and inflaton
perturbations when both are treated linearly%
\footnote{More specifically, we will obtain a Harrison-Zeldovich spectrum
for the scalar metric perturbations with an amplitude which has the right dependence on the parameters of the problem (the Planck mass and the mass of the quadratic inflaton potential). However, our simple calculation does not give the right result for the spectral index: it gives a spectral index whose value is exactly one (rather than slightly smaller than one). In fact, one can explicitly check that the three main approximations that will be employed in this section (namely, neglecting the non-local terms, considering a de Sitter background and computing the quantum correlation function for the inflaton field using the massless approximation) all contribute to a comparable deviation from the exact result for the spectral index, whereas the correct result is obtained when none of the approximations are made. (All this can be checked by proceeding analogously to the calculation in Sec.~8.2.2 of Ref.~\cite{mukhanov05}.)}.
Of course, this fact
ultimately follows from the equivalence between both approaches
established in Sec.~\ref{sec3.3} (as well as in Appendix~\ref{appD}). 
Thus, the purpose of this section
is to illustrate with a simple but relevant example how the
Einstein-Langevin equation can be useful to obtain explicit results
concerning cosmological perturbations.

Let us start by recalling the expression for the $0i$ component of the
Einstein-Langevin equation which was obtained in Sec.~\ref{sec3.2}
working in Fourier space for the spatial components:
\begin{equation}
ik_{i}(\Phi _{k}^{\prime }+\mathcal{H}\Phi _{k})=\frac{\kappa }{2}%
a^{2}\left( \langle ( \delta \hat{\mathcal{T}}_{0}^{i})
_{k}\rangle _{\Phi }+\left( \xi _{0}^{i}\right) _{k}\right)\text{.}
\label{8.110a} 
\end{equation}
The expectation value of the linearized stress tensor operator is
given by Eq.~(\ref{8.79}) and is non-local in the conformal time. In
general, this fact makes it difficult to find an analytic expression
for the solution of the Einstein-Langevin equation.

One possible approach is to realize that there is a certain linear 
combination of the different components of the Einstein-Langevin equation 
for which all the contributions from the non-local terms cancel out, as well 
as those from the stochastic source (a detailed proof and discussion is 
provided in Appendix~\ref{appD}). In that case the equation that one needs 
to solve, Eq.~(\ref{d5}), is a linear second-order ordinary differential 
equation. In fact, this equation has the same form as one often considered 
in standard treatments of linearized cosmological perturbations (see
Eq.~(6.48) in Ref.~\cite{mukhanov92}), and one can take advantage of the 
existing methods and approximation schemes for solving it.
Nevertheless, for illustrative purposes we will not follow this approach in 
this section. We will directly consider Eq.~(\ref{8.110a}), neglect its
non-local part and concentrate on the fluctuating part (neglecting the
non-local term is not necessary, but it simplifies the problem
considerably for a quick calculation).%
\footnote{One can see from our derivations in Sec.~\ref{sec3} that
neglecting the non-local term is equivalent to neglecting the terms
proportional to the metric in Eq.~(\ref{8.71e}).}
%\footnote{One can see from our derivations in Sec.~\ref{sec3} that
%neglecting the non-local term is equivalent to neglecting the terms
%proportional to the metric in Eq.~(\ref{8.71e}), which has been argued
%to be a reasonable approximation IN CERTAIN GAUGES (SUCH AS
%THE COMOVING GAUGE) for small values of the slow-roll
%parameters \cite{lyth90,liddle93}; see also Ref.~\cite{stewart93} for
%a calculation of the higher-order corrections in the slow-roll
%parameters.}).
Eq.~(\ref{8.110a}) then becomes
\begin{equation}
2ik_{i}(\mathcal{H}\Phi _{k}+\Phi _{k}^{\prime })\simeq\frac{8\pi }{m_{p}^{2}}
\left( \xi _{0i}\right) _{k},  \label{8.110}
\end{equation}
from which we can obtain the metric perturbation $\Phi_{k}$ in terms
of the stochastic source $(\xi_{0i})_{k}$.  We need the retarded
propagator for the gravitational potential $\Phi_{k}$, \textit{i.e.},
the required Green function to solve the inhomogeneous first order
differential equation~(\ref{8.110}) with the appropriate boundary
conditions:
\begin{equation}
G_{k}^\mathrm{ret}(\eta ,\eta ^{\prime })=-\frac{i}{2 k_{i}}
\left( \theta (\eta -\eta ^{\prime })\frac{a(\eta ^{\prime })}{a(\eta )}%
+f(\eta ,\eta ^{\prime })\right) ,  \label{8.111}
\end{equation}
where $f(\eta ,\eta ^{\prime })$ is a homogeneous solution related to
the chosen initial conditions. In particular, if we take
$f(\eta,\eta') = -\theta (\eta _{0}-\eta')\ a(\eta')/a(\eta)$, we have
$G_{k}^\mathrm{ret}(\eta ,\eta')=0$ for $\eta \le \eta_0$, which gives
the stochastic evolution of the metric perturbations for $\eta > \eta
_{0}$ due to the effect of the stochastic source after $\eta_{0}$.
The correlation function for the metric perturbations is then given by
the following expression:
\begin{equation}
\langle \Phi _{k}(\eta )\Phi _{k^{\prime }}(\eta ^{\prime })\rangle _{\xi }
\simeq \left(\frac{8\pi }{m_{p}^{2}}\right)^2 \int^{\eta}_{\eta_0}
d\eta_{1} \int^{\eta'}_{\eta_0} d\eta _{2}G_{k}^\mathrm{ret}(\eta,\eta _{1})
G_{k^{\prime }}^\mathrm{ret}(\eta ^{\prime },\eta _{2}) 
 \langle \xi _{k\;0i}(\eta _{1})\xi _{k^{\prime }\;0i}(\eta
_{2})\rangle _{\xi }.  \label{8.112}
\end{equation}
The correlation function for the stochastic source is, in turn,
connected with the stress tensor fluctuations:
\begin{equation}
\langle \xi _{k\;0i}(\eta _{1})\xi _{k'\;0i}(\eta _{2})\rangle _{\xi }=
\frac{1}{2}\langle \{\hat{t}_{0i}^{k}(\eta _{1}),\hat{t}_{0i}^{k'}(\eta
_{2})\}\rangle _{\phi \varphi }
=(2\pi)^3\delta(\vec{k}+\vec{k}') \frac{1}{2}k_{i}k_{i}\phi'(\eta_{1})
\phi ^{\prime }(\eta _{2})G_{k}^{(1)}(\eta _{1},\eta _{2})
\label{8.113}
\end{equation}
where the delta function follows from spatial translational invariance
and $G_{k}^{(1)}(\eta _{1},\eta _{2})=\langle \{\hat{\varphi}_{k}(\eta
_{1}),\hat{\varphi}_{-k}(\eta _{2})\}\rangle $ is the $k$-mode
Hadamard function for a free minimally-coupled scalar field which is
in a state close to the Bunch-Davies vacuum on an almost de Sitter
background.  The so-called \emph{slow-roll} parameters account for the
fact that the background geometry is not exactly that of de Sitter
spacetime, for which $%
a(\eta )=-1/H\eta $ with $-\infty <\eta <0$.

It is also useful to compute the Hadamard function for a massless
field and consider a perturbative expansion in terms of the
dimensionless parameter $m/m_{p}$, for which observations seem to
imply, as will be seen below, a value of the order of $%
10^{-6}$. Thus, we will consider $\bar{G}_{k}^{(1)}(\eta _{1},\eta
_{2})=a(\eta _{1})a(\eta _{2}) G_{k}^{(1)}(\eta _{1},\eta
_{2})=\langle 0|\{%
\hat{y}_{k}(\eta_{1}), \hat{y}_{-k}(\eta _{2})\}|0\rangle$ such that
$\hat{a}%
_k|0\rangle=0$ with $\hat{y}_{k}(\eta)=a(\eta)\hat{\varphi}_{k}(\eta)=
\hat{a%
}_k u_k(\eta)+\hat{a}^\dagger_{-k} u_{-k}^{*}(\eta)$ and $%
u_k(\eta)=(2k)^{-1/2}e^{-ik\eta}(1-i/k\eta)$ corresponding to the
positive frequency $k$-mode for a massless minimally-coupled scalar
field in the Bunch-Davies vacuum state on a de Sitter background
\cite{birrell94}.

The result to lowest order in the mass $m$ of the inflaton field and
the slow-roll parameters is:
\begin{eqnarray}
\langle \Phi _{k}(\eta )\Phi _{k^{\prime }}(\eta ^{\prime })\rangle _{\xi }
&\simeq &\frac{64\pi ^{5}}{m_{p}^{4}}\delta
(\vec{k}+\vec{k}^{\prime })\int_{\eta
_{0}}^{\eta }d\eta _{1}\int_{\eta _{0}}^{\eta ^{\prime }}d\eta _{2}\frac{%
a(\eta _{1})}{a(\eta )}\frac{a(\eta _{2})}{a(\eta ^{\prime })}\dot{\phi}%
(\eta _{1})\dot{\phi}(\eta _{2})\bar{G}_{k}^{(1)}(\eta _{1},\eta _{2})
\nonumber \\
&=&64\pi ^{5}\left( \frac{m}{m_{p}}\right) ^{2}k^{-3}\delta (\vec{k}+\vec{k}%
^{\prime })\int_{k\eta _{0}}^{k\eta }d(k\eta _{1})\int_{k\eta _{0}}^{k\eta
^{\prime }}d(k\eta _{2})\frac{k\eta }{k\eta _{1}}\frac{k\eta ^{\prime }}{%
k\eta _{2}}  \nonumber \\
&&\ \ \ \times\left[\left( 1+\frac{1}{k\eta
_{1}k\eta _{2}}\right)\cos k(\eta _{1}-\eta _{2})
 - \left( \frac{1}{k\eta _{1}}-\frac{1}{k\eta _{2}}\right)
 \sin k(\eta _{1}-\eta _{2})\right]   \nonumber \\
&=&64\pi ^{5}\left( \frac{m}{m_{p}}\right) ^{2}k^{-3}\delta (\vec{k}+\vec{k}%
^{\prime })\left[ \cos k(\eta -\eta ^{\prime })-\frac{1}{k\eta _{0}}\biggl(
k\eta \cos k(\eta -\eta _{0}) \right.   \nonumber \\
&&\left. \quad\quad\quad\quad\quad\quad\quad\quad\quad\quad\quad\quad
+k\eta ^{\prime }\cos k(\eta ^{\prime }-\eta _{0})\biggr) +%
\frac{k\eta k\eta ^{\prime }}{(k\eta _{0})^{2}}\right] \text{,}
\label{8.114}
\end{eqnarray}
where we used the lowest order approximation for $\dot{\phi}(t)$
during slow-roll: $\dot{\phi}(t)\simeq -m_{p}^{2}(m/m_{p})$; overdots
denote here derivatives with respect to the physical time $t$. We
considered the effect of the stochastic source after the conformal
time $\eta _{0}$. Notice that the result (\ref{8.114}) is rather
independent of the value of $\eta _{0}$ provided that it is negative
enough,\emph{\ i.e.}, it corresponds to an early enough initial
time. This weak dependence on the initial conditions is fairly common
in this context and can be qualitatively understood as follows: after
a sufficient amount of time, the accelerated expansion for the
quasi-de Sitter spacetime during inflation effectively erases any
information about the initial conditions, which is redshifted
away. The actual result will, therefore, be very close to that for
$\eta _{0}=-\infty $:
\begin{equation}
\langle \Phi _{k}(\eta )\Phi _{k^{\prime }}(\eta ^{\prime })\rangle _{\xi
}\simeq 8\pi ^{2}\left( \frac{m}{m_{p}}\right) ^{2}k^{-3}(2\pi )^{3}\delta
(\vec{k}%
+\vec{k}^{\prime })\cos k(\eta -\eta ^{\prime })\text{.}  \label{8.115}
\end{equation}

One remark concerning the massless approximation for the computation
of the Hadamard function of the inflaton perturbations is needed. It
is clear from the equation for the scalar field modes that when one
considers scales much smaller than the Compton wavelength of the
inflaton field, \emph{i.e.}, $k/a(\eta )\gg m$, the effect of the mass
term can be neglected. On the other hand, for scales larger than the
Compton wavelength one could object that the mass term should no
longer be negligible. However, it can be argued that the mass term can
also be neglected for large scales provided that the Compton
wavelength is much larger than the horizon (the Hubble radius
$H^{-1}$), \emph{i.e.}, $H\gg m$. The argument goes as follows. For a
massless minimally-coupled scalar field in de Sitter spacetime the
modes become effectively frozen after leaving the horizon: $k/a(\eta
)<H$. On the other hand, for a massive scalar field, the modes decay
approximately like $\exp (-m^{2}\Delta t/3H)$ outside the horizon, but
this decay will not be important if $m^2/H^2$ is small enough. In
particular, if $3 H^2/m^2 \gtrsim 60$, the decay factor $\exp
[-(m^{2}/3H^{2})H\Delta t]$ will not be too different from one for
those modes that left the horizon during the last sixty $e$-folds of
inflation ($H\Delta t=60$ with $\Delta t$ being the time between
horizon exit and the end of inflation), which includes all the
relevant cosmological scales since the scale that left the horizon
sixty $e$-folds before the end of inflation corresponds to the size of
the visible universe at present; any feature with a scale larger than
the visible universe appears to us as observationally
indistinguishable from a homogeneous one. Hence, due to the special
behavior of the modes outside the horizon, even when considering
scales which became much larger than the Compton wavelength before the
end of inflation, $k\exp (-H\Delta t)<m$, it is reasonable to
approximate a massive scalar field with a massless one as long as
$H\gg m$, which happens to be the case in most slow-roll inflationary
models and in particular for the simple example considered in this
section \cite{linde90}.

Let us consider the cosmological implications which can be extracted
from Eq.~(\ref{8.115}), especially those related to large-scale
gravitational fluctuations. These fluctuations are believed to play a
crucial role in the generation of the large-scale structure and matter
distribution observed in our present universe
\cite{padmanabhan93}. They are also closely connected with the
anisotropies in the CMB radiation, which decoupled from matter about
$4 \times 10^{5}$ years after the Big Bang and provides us with very
valuable information about the early universe
\cite{kolb90,mukhanov05}.

{}From the analysis of our final result in Eq.\ (\ref{8.115}) two main
well-known facts can be concluded. First, an almost
Harrison-Zeldovich scale-invariant spectrum is obtained for large
scales.  Indeed, for scales clearly outside the horizon at the times
$\eta $ and $%
\eta ^{\prime }$, \emph{i.e.}, $k\eta ,\,k\eta ^{\prime }\ll 1$, the
right-hand side of Eq.~(\ref{8.115}) becomes proportional to
$k^{-3}\delta (%
\vec{k}+\vec{k}^{\prime })$ with negligible extra dependence on $k$,
$\eta $ and $\eta ^{\prime }$.  Second, since we get $\langle \Phi
_{k}(\eta )\Phi _{k^{\prime }}(\eta ^{\prime })\rangle _{\xi }\propto
(m/m_{p})^{2}$ in agreement with the usual results \cite
{mukhanov92,tanaka98,mukhanov05}, the small value of the CMB
anisotropies first detected by COBE imposes a severe bound on the
gravitational fluctuations, characterized by $\langle \Phi _{k}(\eta
)\Phi _{k^{\prime }}(\eta ^{\prime })\rangle _{\xi }$, which implies
the following restriction (fine tuning) for the inflaton mass:
$m/m_{p}\sim 10^{-6}$.

Some comments on the mechanisms considered in earlier related work
\cite{calzetta95a,matacz97a,matacz97b,calzetta97a} which allowed a
significant relaxation on the fine tuning of that kind of parameters
are in order here.  In those studies either a self-interacting scalar
field or a scalar field interacting non-linearly with other fields
were considered. The modes of the inflaton field corresponding to
scales of cosmological interest were regarded as an open quantum
system with the environment constituted either by the short-wavelength
modes in the case of self-interaction or else by other fields
interacting with the inflaton field. Therefore, one can introduce a
stochastic description based on a Langevin equation, as explained in
Ref.~\cite{calzetta03a}, to study the dynamics of the inflaton field
modes. In fact, Langevin-type equations or related stochastic tools
were employed in the references cited above.
Furthermore, it was shown in Ref.~\cite{calzetta03a} that the validity
of the results, such as the correlation functions, obtained by those
methods is independent of the existence of enough decoherence to
guarantee the presence of a semiclassical regime for the system
dynamics. However, it was also shown that the two-point quantum
correlation function for the system had two separate contributions
(see Eq.~(4.9) in Ref.~\cite{calzetta03a}): one related to the
dispersion of the system's initial state and another one which was
proportional to the noise kernel and accounted for the fluctuations of
the system induced by the interaction with the environment. For
natural states in de Sitter spacetime such as the Bunch-Davies vacuum
(in fact, any reasonable state in de Sitter space tends asymptotically
to it \cite{anderson00}) the dispersion is proportional to
$H^{2}$. This contribution is actually several orders of magnitude
larger than that coming from the term proportional to the noise kernel
for the situations considered in
Refs.~\cite{calzetta95a,matacz97a,matacz97b,calzetta97a}. Thus, the
two-point quantum correlation function for the inflaton perturbations
is dominated by the contribution connected to the dispersion of the
initial state. This point has been confirmed by a detailed analysis in
Ref.~\cite{lombardo05}.  Moreover, this contribution essentially
coincides for the cases of an interacting and a free scalar field. The
latter is the case being considered throughout this paper and exhibits
no noise term for the inflaton dynamics because there is no
environment for the inflaton perturbations (this should not be
confused with the noise kernel for the fluctuations of the metric
perturbations induced by the quantum fluctuations of the
inflaton). One could try to choose the initial state, as argued in
Ref.~\cite{calzetta97a}, so that the contribution from the dispersion
of the initial conditions were smaller than the fluctuations induced
on the modes of the inflaton field, but that would require a great
amount of fine-tuning for the initial quantum state of each mode,
which would become highly unstable due to the large dispersion in
momentum implied by Heisenberg's uncertainty principle \cite{tanaka98}
and quickly tend to the de Sitter invariant Bunch-Davies vacuum; or
even have no inflation at all due to the back reaction on the
evolution of the background geometry generated by such a highly
excited state.

\section{Discussion}

\label{sec5}

In this paper we have studied linearized metric perturbations around a
Robertson-Walker background interacting with a quantum scalar field
and we have shown that, when linearizing the perturbations of the
scalar field around its background configuration, the
Einstein-Langevin equation yields a result for the correlation
function of the metric perturbations equivalent to that obtained in
the usual approach based on the linearization and quantization of both
the metric perturbations and the perturbations of the scalar field
around its expectation value. Although for the sake of concreteness we
have mostly concentrated on the case of a spatially flat
Robertson-Walker metric and a minimally-coupled scalar field with a
quadratic potential, the main result can be generalized rather
straightforwardly to Robertson-Walker metrics with non-flat
homogeneous spatial sections, as well as to general potentials for the
scalar field and arbitrary coupling to the spacetime
curvature. Considering Robertson-Walker metrics with homogeneous
spatial sections of positive or negative curvature would imply,
respectively, the use of three-dimensional spherical or hyperbolic
harmonics
%***** check this *****
rather than simple Fourier transforms for the spatial coordinates, but
that would not substantially change the procedures and the main
conclusions since the basic properties of Fourier transforms employed
in the text have analogous counterparts for these harmonics
\cite{bander66,bunch78b,mottola85}.  On the other hand, the use of a
general potential should not imply major differences since after all
we would linearize with respect to the scalar field perturbations
around the background configuration.

In addition, we also provided in Sec.~\ref{sec4} a particular example
illustrating how the Einstein-Langevin equation can be used in
practice to compute the correlation function at large scales for
scalar metric perturbations in cosmological inflationary models. In
doing so we made use of slightly oversimplified approximations,
namely, use of a de Sitter background geometry, calculation of the
Hadamard function for a massless scalar field and neglecting a
non-local term, because the result has already been computed in a
number of references, see for instance
Refs.~\cite{kodama84,mukhanov92,linde90,mukhanov05}, and our primary
concern was simply to show how the Einstein-Langevin equation can be
used to obtain an explicit result for cosmological perturbations.

Throughout the article we have concentrated on scalar-type metric
perturbations. The reason for this is that, when linearizing with
respect to both the metric perturbations and the scalar field
perturbations, the vectorial and tensorial metric perturbations
decouple from the matter scalar field. In this case the metric
perturbations do not constitute a true open system since the dynamics
of the scalar and vectorial perturbations is completely constrained by
the temporal components of the Einstein equation; in fact, the
vectorial ones actually turn out to vanish and the scalar ones cannot
be regarded as a degree of freedom independent of the scalar field
perturbations. Moreover, the only true dynamical degrees of freedom,
the two tensorial ones, do not couple to the matter field. On the
other hand, an even more interesting situation corresponds to the case
in which the scalar field is treated exactly, at least for quadratic
potentials. Then the scalar field also couples to the metric
perturbations of tensorial type and the metric perturbations become a
true open system with the scalar field corresponding to the
environment.

The main features that would characterize an exact treatment of the
scalar field perturbations interacting with the metric perturbations
around a Robertson-Walker background as compared to the case addressed
in this paper are the following. First, the three types of metric
perturbations couple to the perturbations of the scalar field, as
already mentioned above. Second, the corresponding Einstein-Langevin
equation for the linear metric perturbations will explicitly couple
the scalar and tensorial metric perturbations. Furthermore, although
the Fourier modes (with respect to the spatial coordinates) for the
metric perturbations will still decouple in the Einstein-Langevin
equation, any given mode of the noise and dissipation kernels will get
contributions from an infinite number of Fourier modes of the scalar
field perturbations (see Ref.~\cite{roura99b} for an explicit calculation
of the noise kernel for a massless and minimally-coupled scalar field
in de Sitter). This fact will imply, in addition, the need to
properly renormalize the ultraviolet divergences arising in the
dissipation kernel, which actually correspond to the divergences
associated with the expectation value of the stress tensor operator of
the quantum matter field evolving on the perturbed geometry.

The importance of considering corrections due to one-loop
contributions from scalar field perturbations, beyond the tree level
of the linear cosmological perturbation theory, has recently been
emphasized \cite{weinberg05,weinberg06}. In the present context this
means treating the scalar field perturbations exactly in the
Einstein-Langevin equation.  Furthermore, in Ref.~\cite{calzetta03a}
it was explained how a stochastic description based on a Langevin-type
equation could be introduced to gain information on fully quantum
properties of simple linear open systems. In a forthcoming paper
\cite{roura03b} it will be shown that, by carefully dealing with the
gauge freedom and the consequent dynamical constraints, the previous
result can be extended to the case of $N$ free quantum matter fields
weakly interacting with the metric perturbations around a given
background (here weakly interacting means that the gravitational
coupling constant times the number of fields remains constant in the
limit of large $N$). In particular, the correlation functions for the
metric perturbations obtained using the Einstein-Langevin equation are
equivalent to the leading order contribution in the large $N$ limit to
the correlation functions that would follow from a purely quantum
field theory calculation. This will generalize the results already
obtained on a Minkowski background \cite{hu04b,hu04c}.  These results
have important implications on the use of the Einstein-Langevin
equation to address situations in which the background configuration
for the scalar field vanishes, so that linearization around such a
configuration is no longer possible. This includes not only the case
of a Minkowski background spacetime, but also the remarkably
interesting case of inflationary models driven by the vacuum
polarization of a large number of conformal fields with vanishing
expectation value \cite{starobinsky80,vilenkin85,hawking01}, where the
usual approaches based on the linearization of both the metric
perturbations and the scalar field perturbations and their subsequent
quantization can no longer be applied.

\begin{acknowledgments}
We are grateful to Daniel Arteaga, Esteban Calzetta, Antonio Campos,
Jaume Garriga, Bei-Lok Hu, Kei-ichi Maeda, Rosario Mart\'{\i}n and
Yuko Urakawa for many interesting discussions. This work has been
partially supported by the Research Projects MEC FPA-2004-04582 and
DURSI 2005SGR00082.  During the last stages of this project A.~R.\ has
also been supported by LDRD funds from Los Alamos National Laboratory.
\end{acknowledgments}

\appendix

\section{Definition and basic properties of Gaussian states}

\label{app0}

In this appendix we summarize the definition and basic properties of
Gaussian pure states. In contrast to the rest of the paper, $\Psi$
will be used throughout this appendix to denote either the state of a
field or its wave functional in the Schr\"odinger picture rather than
a gauge invariant variable for scalar-type metric perturbations.

A pure state is called \emph{Gaussian} if its wave functional in the
Schr\"odinger picture is a Gaussian functional:
\begin{equation}
\Psi\left[\phi(\vec{x})\right] \propto \exp \left[ -\int d^3x\, d^3x'
\phi(\vec{x})A(\vec{x},\vec{x}')\phi(\vec{x}') + \int d^3x B(\vec{x})
\phi(\vec{x}) \right]
\label{a1},
\end{equation}
with a suitable normalization constant.  The fundamental property of
Gaussian states is the fact that the cumulants of order higher than
two associated with the quantum expectation values of products of the
field operator $\hat{\phi}$ vanish, \emph{i.e.},
\begin{equation}
\left. \frac{1}{i^n}\frac{\delta}{\delta j(\vec{x}_1)} \ldots \frac{\delta}
{\delta j(\vec{x}_n)} \ln \left\langle \exp^{i \int d^3x j(\vec{x})
\hat{\phi}(\vec{x})} \right\rangle_\Psi \right|_{j=0} =0
\label{a2},
\end{equation}
for $n \geq 3$ and where we introduced the notation
$\langle\hat{O}\rangle_{\Psi} \equiv
\langle\Psi|\hat{O}|\Psi\rangle$. This implies that the connected part
of any quantum correlation function $\left\langle \Psi \right|
\hat{\phi}(\vec{x}_n)\ldots\hat{\phi}(\vec{x}_1)
\left|\Psi\right\rangle$ with $n \geq 3$ vanishes or, equivalently,
that any quantum correlation function $\left\langle \Psi \right|
\hat{\phi}(\vec{x}_n)\ldots\hat{\phi}(\vec{x}_1)
\left|\Psi\right\rangle$ can be written as a linear combination of
products involving the expectation values $\left\langle \Psi \right|
\hat{\phi}(\vec{x}_i) \left|\Psi\right\rangle$ and two-point functions
$\left\langle \Psi \right| \hat{\phi}(\vec{x}_j) \hat{\phi}(\vec{x}_k)
\left|\Psi\right\rangle$.  Furthermore, if the Hamiltonian of the
field under consideration is quadratic, this property can be
generalized for different times to quantum correlation functions in
the Heisenberg picture of the form
$\sideset{_H}{}{\operatorname{\left\langle \Psi \right|}}
\hat{\phi}(t_n,\vec{x}_n)\ldots\hat{\phi}(t_1,\vec{x}_1)
\left|\Psi\right\rangle_H$, which follows from Wick's theorem
\cite{itzykson80}.

Finally, given a field operator $\hat{\phi}$ and a Gaussian state
$|\Psi\rangle$ with non-vanishing expectation value
$\langle\hat{\phi}\rangle_{\Psi}$, it is always possible to introduce
a new field $\hat{\varphi}=\hat{\phi} -
\langle\hat{\phi}\rangle_{\Psi}$ so that the wave functional
$\tilde{\Psi}[\varphi]$ for the state $|\Psi\rangle$ in the basis
associated with the field $\hat{\varphi}$ becomes a Gaussian
functional with vanishing expectation value and independent of the
expectation value $\langle\hat{\phi}\rangle_{\Psi}$. This can be
immediately seen by rewriting the expression for the wave functional in
Eq.~(\ref{a1}) as
\begin{equation}
\Psi\left[\phi(\vec{x})\right] \propto \exp \left[ -\int d^3x
d^3x'\left(\phi(\vec{x})-\langle\hat{\phi}(\vec{x})\rangle_{\Psi}\right)
A(\vec{x},\vec{x}') \left(\phi(\vec{x}') -
\langle\hat{\phi}(\vec{x})\rangle_{\Psi} \right) \right]
\label{a6},
\end{equation}
and then change to the basis associated with the field $\hat{\varphi}$:
\begin{equation}
\tilde{\Psi}\left[\varphi(\vec{x})\right] \propto \exp \left[ -\int d^3x
d^3x' \varphi(\vec{x}) A(\vec{x},\vec{x}')\varphi(\vec{x}') \right]
\label{a7}.
\end{equation}
It is precisely in this sense that the state for the inflaton field
perturbations $\hat{\varphi}$ introduced in Sec.~\ref{sec2.2} follows
immediately from the state of the inflaton field $\hat{\phi}$.

It should be emphasized that any of the vacuum states commonly
considered for free fields in curved spacetimes are Gaussian
states. Furthermore, as stated in Sec.~\ref{sec2.2}, in this paper we
concentrate on states which are invariant under the symmetries of the
Robertson-Walker metric. In particular, for the case of
Robertson-Walker metrics with flat spatial sections this implies
$\langle\hat{\phi}\rangle_{\Psi}(t,\vec{x}) =
\langle\hat{\phi}\rangle_{\Psi}(t)$ and $A(\vec{x},\vec{x}') =
A(|\vec{x}-\vec{x}'|)$.

\section{Conventions and notation for the linearized stress tensor}

\label{appA}

Here we explain the notation concerning the linearized stress tensor
employed in this article and clarify some related subtle points.

Let us begin with the objects which appear in the Einstein-Langevin
equation (\ref{7.7}). The tensors $G^{(1)}_{ab}$ and $T^{(1)}_{ab}$
correspond to the terms proportional to the metric perturbations in
the perturbed version of the background (unperturbed) objects
$G^{(0)}_{ab}$ and $T^{(0)}_{ab}$. The indices of $G^{(1)}_{ab}$ can
be raised using the background metric [from now on everything that
will be said for $G^{(1)}_{ab}$ applies exactly in the same way to
$T^{(1)}_{ab}$]. On the other hand, one could perturb the background
object $G^{(0)\,ab}$ with the indices already raised, and reach a
different result for $G^{(1)\,ab}$: it would differ by the terms
$G^{(0)\,cd}h^{a}_{c}+G^{(0)\,ac}h^{b}_{c}$. Our notation (and those
commonly employed) is ambiguous in the sense that it does not
distinguish between both possibilities. In order to remove such an
ambiguity it is necessary (and sufficient) to specify \emph{a priori}
which objects are going to be perturbed. In particular, in
Sec.~\ref{sec2} it is $G^{(0)}_{ab}$ and $T^{(0)}_{ab}$ that are
perturbed (of course, one can then freely use the background metric to
raise and lower indices), whereas it is $G^{(0)\,b}_{\ \ \ a}$ and $
T^{(0)\,b}_{\ \ \ a}$ that will be perturbed in Secs.~\ref{sec3} and
\ref{sec4}, and finally $G^{(0)\,ab}$ and $T^{(0)\,ab}$ in
Appendix~\ref{appC}. Fortunately, this will not change the form of the
Einstein-Langevin equation since the terms corresponding to the
difference between both choices for the Einstein tensor and the stress
tensor cancel out because their background counterparts satisfy the
Einstein equation $G^{(0)}_{ab}=\kappa T^{(0)}_{ab}$. Hence, if one
deals with equations involving tensorial objects rather than with
isolated objects, everything is independent of the particular choice
provided that the same choice is made for all the objects in the
equation.

The previous ambiguity does not affect the stochastic source of the
Einstein-Langevin equation, which is completely defined on the
background spacetime (the noise kernel is evaluated on the background
geometry). Furthermore, the argument given in Sec.~\ref{sec3} to show
that $\Psi=\Phi$ is not affected by such an ambiguity either. The
reason is that the ambiguous extra terms for $\delta\mathcal{T}_{ij}$
with $i \neq j$ vanish because both $\langle \hat{T}^{(0)}
_{\mu\nu}\rangle_\mathrm{ren}$ (or $\mathcal{T}_{\mu\nu}$) and the
scalar metric perturbations in the longitudinal gauge are diagonal.
Similarly, each one of the terms appearing in the $0i$ component of
the Einstein-Langevin and the quantum version of the linearized
Einstein equations considered in detail in Secs.~\ref{sec3.2} and
\ref{sec3.3}, respectively, do not suffer from the ambiguity either.
%This is because the metric perturbations are diagonal and the $0i$
%components of the
%background expectation value, $\left\langle \hat{T}^{(0)}
%_{0i}\right\rangle_\mathrm{ren}$, vanish.

The tensors $G^{(1)}_{ab}$ and $T^{(1)}_{ab}$, which are discussed
above, result from linearizing just the metric perturbations. On the
other hand, in the usual treatment of cosmological perturbations in
inflationary models not only the metric perturbations, but also the
inflaton perturbations are simultaneously linearized. Therefore, we
introduced the notation $\mathcal{T}_{ab}$ for the contribution to the
expectation value of the background stress tensor $\langle
\hat{T}^{(0)}_{ab} \rangle_\mathrm{ren}$ which is quadratic in the
background solution $\phi(\eta)$ and independent of the inflaton
perturbations, \emph{i.e.}, the first term on the right-hand side of
Eq.~(\ref{8.4}): $\mathcal{T}_{ab}=\langle \hat{T}^{(0)}_{ab}\rangle
_{\phi\phi }$.  As we already explained in Sec.~\ref{sec2.2}, the
expectation value $\langle \hat{T}^{(0)}_{ab} \rangle_\mathrm{ren}$
has also a contribution which is quadratic in the inflaton
perturbations, but it is neglected when linearizing in those. We also
introduced the notation $\delta\hat{\mathcal{T}}_{ab}$ for the
perturbed stress tensor operator obtained when linearizing with
respect to \emph{both} the metric perturbations and the inflaton
perturbations. Its expectation value $\langle
\delta\hat{\mathcal{T}}_{ab}[g+h]\rangle$ is, thus, equivalent to
linearizing also with respect to the inflaton perturbations the
expectation value $\langle
\hat{T}^{(1)}_{ab}[g+h]\rangle_\mathrm{ren}$. Similarly, the operator
$\delta\hat{\mathcal{T}}_{ab}[g+h]$ corresponds to linearizing in the
inflaton perturbations the expression $\hat{T}^{(0)} _{ab}-\langle
\hat{T}^{(0)}_{ab} \rangle+\hat{T}^{(1)}_{ab}$. This expression, which
appears (after raising one index) in Eq.~(\ref{8.44}), might seem a
bit awkward at that point, but this is just because the notation
generally employed for the Einstein-Langevin equation, where one
linearizes only with respect to the metric perturbations, is no longer
the most natural when one also linearizes with respect to the inflaton
perturbations. The expression is appropriate either when quantizing
both the metric perturbations and the scalar field perturbations or
when considering a stochastic version of it, namely the
Einstein-Langevin equation. In the latter case one takes the
expectation value of $\hat{T}^{(1)}_{ab}$ plus a stochastic source
that accounts for the quantum fluctuations of the operator
$\hat{t}_{ab}=\hat{T}^{(0)}_{ab} -\langle \hat{T}^{(0)}_{ab} \rangle$,
whose expectation value vanishes. Finally, the notation
$\delta\hat{t}_{ab}$ is used for the result of linearizing the
operator $\hat{t}_{ab}$ with respect to the inflaton perturbations,
which coincides with $\delta\hat{\mathcal{T}}_{ab}$ evaluated on the
unperturbed metric.

\section{Functional approach to the Einstein-Langevin equation
and alternative derivation of Eq.~(\ref{8.79})}

\label{appB}

The Einstein-Langevin equation for metric perturbations around a given
background and interacting with quantum matter fields has been
formally derived using functional methods
\cite{calzetta94,hu95a,hu95b,campos96,calzetta97c,martin99b,martin99c}.
This was achieved by regarding the metric perturbations as an open
quantum system with the environment corresponding to the quantum
matter fields, and using the influence functional formalism for open
quantum systems introduced by Feynman and Vernon
\cite{feynman63,feynman65}. In this appendix we briefly review some
basic aspects of the functional approach to the Einstein-Langevin
equation and explain how an alternative derivation of Eq.~(\ref{8.79})
for the expectation value of the linearized stress tensor operator
evaluated on the perturbed metric can be obtained.

When considering derivations of the Einstein-Langevin equation using
functional methods, one begins by computing the influence functional
for the metric perturbations by integrating out the quantum matter
fields (we will only consider free fields) as follows:
\begin{equation}
e^{iS_\mathrm{IF}\left[ h,h^{\prime }\right]}
=\int d\varphi _\mathrm{f} d\varphi _\mathrm{i} d\varphi_\mathrm{i}^{\prime}
\int_{\varphi (t_\mathrm{i})=\varphi _\mathrm{i}}^{\varphi (t_\mathrm{f})
=\varphi_\mathrm{f}}\mathcal{D}\varphi \int_{\varphi ^{\prime}(t_\mathrm{i})
=\varphi _\mathrm{i}^{\prime}}^{\varphi ^{\prime }(t_\mathrm{f})
=\varphi _\mathrm{f}} \mathcal{D}\varphi ^{\prime }e^{iS\left[ \varphi ,g+h\right]
-iS[\varphi',g+h^{\prime}]}\rho \left[ \varphi _\mathrm{i},\varphi _\mathrm{i}
^{\prime};t_\mathrm{i}\right) \text{,}  \label{7.8}
\end{equation}
where $\rho \left[ \varphi _\mathrm{i},\varphi _\mathrm{i}^{\prime };
  t_\mathrm{i}\right) $ is the density matrix for the initial state of
the matter field, $\varphi $, which is assumed to be initially
uncorrelated with the metric perturbations (moreover, asymptotic
initial conditions with $t_\mathrm{i}\rightarrow -\infty $ are usually
considered) and $S\left[ \varphi ,g+h\right] $ is the action for the
matter field evolving on a spacetime with metric $g_{ab}+h_{ab}$.
Furthermore, only terms up to quadratic order in the metric
perturbations $h_{ab}$ around the fixed background metric $g_{ab}$
will be considered. In that case the action for the matter field can
be written as
\begin{eqnarray}
S\left[ \varphi ,g+h\right]  &=&\frac{1}{2}\int d^{4}x\sqrt{-g\left( x\right)
}h_{ab}\left( x\right) T^{ab}\left[ \varphi ,g_{ab};x\right)
\nonumber \\
&&+\frac{1}{4}\int d^{4}x\sqrt{-g\left( x\right) }\int d^{4}y\sqrt{-g\left(
y\right) }h_{ab}\left( x\right) h_{cd}\left( y\right)   \nonumber \\
&&\ \ \ \ \ \ \ \frac{1}{\sqrt{-g(x)}\sqrt{-g(y)}}\frac{\delta ( \sqrt{%
-g\left( x\right) }T^{ab}\left[ \varphi ,g_{ab};x\right)) }{\delta
g_{cd}\left( y\right) } +O\left( h_{ab}^{3}\right) \text{,}
\label{7.9}
\end{eqnarray}
where $T^{ab}\left[ \varphi ,g_{ab}^{\prime };x\right) =2\left( -g\left(
x\right) \right) ^{-1/2}\delta S\left[ \varphi ,g\right] /\delta g_{ab}\left(
x\right) $ corresponds to the stress tensor for the matter field, whose
functional derivative $\left( -g\left( y\right) \right) ^{-1/2}\delta
T^{ab}\left[ \varphi ,g_{ab};x\right) /\delta g_{cd}\left( y\right) $ is a
local object, \emph{i.e.}, proportional to the covariant delta function $\left(
-g\left( y\right) \right) ^{-1/2}\delta ^{(4)}\left( x-y\right) $. The
influence action to quadratic order in the metric perturbations, which can be
obtained by integrating out the matter field $\varphi$ as explained in
Refs.~\cite{martin99b,martin99c}, exhibits a structure analogous to
that of a linear open quantum system:
\begin{equation}
S_\mathrm{IF}\left[ \Sigma ,\Delta \right] =Z\cdot \Delta +\Delta \cdot (H+K)\cdot
\Sigma
+\frac{i}{8}\Delta \cdot N\cdot \Delta \text{,}  \label{7.10}
\end{equation}
where $A\cdot B$ denotes $\int d^{4}x\sqrt{-g\left( x\right)}
A_{ab}\left( x\right) B^{ab}\left( x\right) $ and we introduced the
average and difference variables $\Sigma _{ab}=(h_{ab}+h_{ab}^{\prime
})/2$ and $\Delta _{ab}=h_{ab}^{\prime }-h_{ab}$. The expressions for
the kernels are the following:
\begin{eqnarray}
Z^{ab}\left( x\right)  &=&-\frac{1}{2}\langle \hat{T}^{ab}\left[ \hat{%
\varphi},g_{ab};x\right) \rangle \text{,}  \label{7.11} \\
H^{abcd}\left( x,y\right)  &=& \frac{1}{4}\mathrm{Im} \langle T^{*}\hat{T}%
^{ab}\left[ \hat{\varphi},g;x\right) \hat{T}^{cd}\left[ \hat{\varphi}%
,g;y\right) \rangle -\frac{i}{8}\langle [ \hat{T}^{ab}\left[
\hat{\varphi},g;x\right) ,\hat{T}^{cd}\left[ \hat{\varphi},g;y\right)
] \rangle   \text{,}  \label{7.12} \\
K^{abcd}(x,y) &=&  \frac{-1}{2\sqrt{-g\left( x\right)}
\sqrt{-g\left( y\right)}} \frac{\delta (\sqrt{-g\left( x\right) }
\langle \hat{T}^{ab}\left[ \hat{\varphi},g_{ab};x\right) \rangle
) }{\delta g_{cd}\left( y\right) }  ,  \label{7.12b}\\
N^{abcd}\left( x,y\right)  &=&\frac{1}{2}\left\langle \left\{ \hat{t}%
^{ab}\left[ \hat{\varphi},g;x\right) ,\hat{t}^{cd}\left[ \hat{\varphi}%
,g;y\right) \right\} \right\rangle \text{,}  \label{7.13}
\end{eqnarray}
where the functional derivative in Eq.~(\ref{7.12b}) should be
understood to account only for the explicit dependence on the metric,
whereas the implicit dependence through the field operator
$\hat{\varphi}[g]$ is not included.  The notation $T^{*}$ appearing in
Eq.~(\ref{7.12}) means that the matter field operators must be
temporally ordered before applying any derivatives acting on
them. Thus, we have, for instance, $\left\langle T^{*}\nabla
  _{a}^{x}\hat{\varphi}(x)\nabla _{b}^{y}%
  \hat{\varphi}(y)\right\rangle =$ $\nabla _{a}^{x}\nabla
_{b}^{y}\left\langle T\hat{\varphi}(x)\hat{\varphi}(y)\right\rangle
$. Note that although the background geometry is non-trivial in
general, the notion of temporal ordering is well defined because we
are restricting the possible background geometries to globally
hyperbolic spacetimes, which are time orientable; moreover, the
microcausality condition of the quantum field theory for the matter
fields under consideration guarantees that $\langle [
\hat{O}_{1}(x),\hat{O}_{2}(y)] \rangle =0$ if $\hat{O}_{1}(x)$ and
$\hat{O}_{2}(y)$ are local operators and $x$ and $y$ are spacelike
separated points. It should also be noted that the first term on the
right-hand side of Eq.~(\ref {7.12}) is symmetric under interchange of
$x$ and $y$, whereas the second one is completely antisymmetric. On
the other hand, the term on the right-hand side of Eq.~(\ref{7.12b})
is local and symmetric under interchange of $x$ and $y$.

The noise kernel $N^{abcd}(x,y)$ requires no renormalization, as
explained in Sec.~\ref{sec2.1}, whereas the kernels $Z^{ab}(x)$,
$H^{abcd}(x,y)$ and $K^{abcd}(x,y)$ contain divergences [some
regularization procedure is implicitly understood in
Eqs.~(\ref{7.11})-(\ref{7.13})] that can be canceled out by adding
suitable counterterms, quadratic in the curvature, to the bare
gravitational action. These are precisely the same counterterms which
are introduced in semiclassical gravity so that, when functionally
differentiating with respect to the metric, they cancel the
divergences from the expectation value of the stress tensor. This fact
should not be surprising at all since the kernel $Z^{ab}(x)$
corresponds to the expectation value of the stress tensor operator on
the background metric and the kernels $H^{abcd}(x,y)$ and
$K^{abcd}(x,y)$ are closely related to the expectation value of the
stress tensor operator on the perturbed metric, as follows
straightforwardly from the following relation, valid up to linear
order in $h_{ab}$:
\begin{equation}
\langle \hat{T}^{ab}\left[ g+h;x\right) \rangle =\frac{2}
{\sqrt{-(g+h)(x)}}\left. \frac{\delta S_\mathrm{IF}\left[ h,h'\right] }{\delta
h_{ab}(x)}\right|_{h'=h}= -2Z^{ab}(x)-2(H\cdot h)^{ab}(x)-2(M\cdot h)^{ab}(x)
\text{,}  \label{7.14}
\end{equation}
where we introduced the kernel $M^{abcd}(x,y)$ defined as follows:
\begin{equation}
M^{abcd}(x,y) \equiv
 \frac{-1}{2\sqrt{-g\left( y\right)}} \frac{\delta (
\langle \hat{T}^{ab}\left[ \hat{\varphi},g_{ab};x\right) \rangle
) }{\delta g_{cd}\left( y\right) }  , \label{7.14b}
\end{equation}
which results from adding to the kernel $K^{abcd}(x,y)$ the term
$-(-g(x)) ^{-1/2} (\delta\sqrt{-g(x)}/\delta g_{cd}(y)) Z^{ab}$ coming
from the contribution to the factor $2 [-(g+h)(x)]^{-1/2}$ that is
linear in $h_{ab}$. When the counterterms introduced in the bare
gravitational action are included in the influence action, so that the
divergences cancel out and the bare kernels $H^{abcd}(x,y)$ and
$M^{abcd}(x,y)$ get renormalized, Eq.~(\ref{7.14}) becomes
\begin{equation}
\langle \hat{T}^{ab}\left[ g+h;x\right) \rangle _\mathrm{ren}=\frac{2}
{\sqrt{-(g+h)(x)}}\left. \frac{\delta S_\mathrm{IF}^\mathrm{(ren)}\left[h,h'\right]} {\delta
h_{ab}\left( x\right) }\right| _{h^{\prime }=h}=\langle \hat{T}^{ab}
\left[ g;x\right) \rangle _\mathrm{ren}-2\left(H_\mathrm{ren}\cdot h\right)
^{ab}(x)-2\left(M_\mathrm{ren}\cdot h\right)^{ab}(x)  \text{,} \label{7.15}
\end{equation}
which can be rewritten as
\begin{equation}
\langle \hat{T}^{(1)\,ab}\left[ g+h;x\right) \rangle
_\mathrm{ren}=-2\left(H_\mathrm{ren}\cdot h\right)^{ab}(x)
-2\left(M_\mathrm{ren}\cdot h\right)^{ab}(x) , \label{7.15b}
\end{equation}
where, as mentioned above, the functional derivative appearing in the
kernel $M^{abcd}(x,y)$ should be understood to account only for the
explicit dependence of the stress tensor on the metric, whereas the
implicit dependence through the field operator $\hat{\varphi}[g]$ is
entirely contained in the first term on the right-hand side of
Eq.~(\ref{7.15b}).

Taking into account the previous results, the Einstein-Langevin
equation can then be obtained from the CTP effective action for the
metric perturbations by using a formal trick described below. Such a
CTP effective action for the metric perturbations has the following
form at tree level (note, however, that the matter fields, which have
already been integrated out, were treated beyond the tree level):
\begin{equation}
\Gamma _\mathrm{CTP}^{(0)}\left[ h,h^{\prime }\right] =S_\mathrm{g}\left[ h\right]
-S_\mathrm{g}\left[ h^{\prime }\right] +S_\mathrm{IF}^\mathrm{(ren)}\left[ h,h^{\prime }\right]
+O\left(h_{ab}^{3}\right) \text{,}  \label{7.16}
\end{equation}
where $S_\mathrm{g}\left[ h\right] $ is the Einstein-Hilbert action
$\int d^{4}x%
\sqrt{-g}R$ up to quadratic order in the metric perturbations, and the
finite parts of the local counterterms have been included in
$S_\mathrm{IF}^\mathrm{(ren)}$. On the other hand, using the following
mathematical identity for the imaginary part of the influence action:
\begin{equation}
e^{-\mathrm{Im} S_\mathrm{IF}}=
e^{-\frac{1}{8}\Delta \cdot N\cdot \Delta }=\det \left( 2\pi N\right) ^{-%
\frac{1}{2}}\int \mathcal{D}\xi e^{-\frac{1}{2}\xi \cdot N^{-1}\cdot \xi
}e^{-\frac{i}{2}\xi \cdot \Delta }\text{,}  \label{7.17}
\end{equation}
and interpreting $\xi ^{ab}$ as a stochastic source with vanishing
expectation value and correlation function $\left\langle \xi ^{ab}(x)\xi
^{cd}(y)\right\rangle _{\xi }=N^{abcd}(x,y)$, one can define a
\emph{stochastic} effective action,
\begin{equation}
\Gamma _\mathrm{stoch}\left[ \Sigma ,\Delta \right] =S_\mathrm{g}\left[ h\right]
-S_\mathrm{g}\left[ h^{\prime }\right] +\Delta \cdot (H_\mathrm{ren}+M_\mathrm{ren})\cdot \Sigma
-\frac{1}{2}\xi \cdot \Delta \text{,}  \label{7.18}
\end{equation}
such that $\langle\exp(i\Gamma_\mathrm{stoch})\rangle_{\xi}=
\exp(i\Gamma_\mathrm{CTP}^{(0)})$.  The Einstein-Langevin equation can
be immediately obtained by functionally differentiating with respect
to the metric perturbation $h_{ab}$ and letting $h_{ab}'=h_{ab}$
afterwards:
\begin{equation}
0= \left.\frac{1}{\sqrt{-(g+h)(x)}} \frac{\delta \Gamma _\mathrm{stoch}}{\delta
h_{ab}(x)}\right|_{h^\prime=h} =-\frac{1}{2\kappa }
G^{(1)\,ab}\left[ g+h;x\right) -\left( H_\mathrm{ren}\cdot h\right)
^{ab}\left(x\right) -\left(M_\mathrm{ren}\cdot h\right)^{ab}\left(x\right)+\frac{1}
{2}\xi^{ab}\left( x\right) \text{.}  \label{7.19}
\end{equation}
It is worth discussing the issue of causality in Eq.~(\ref{7.19}),
which basically amounts to considering the second term on the
right-hand side of the equation since the remaining terms are
local. The right-hand side of Eq.~(\ref{7.12}) can be formally
rewritten as
\begin{equation}
- \frac{i}{4}\left\langle \left[ \hat{t}^{ab}\left[ \hat{\varphi} ,g;x\right)
,\hat{t}^{cd}\left[ \hat{\varphi} ,g;y\right) \right] \right\rangle \theta ^{*}
\left( \eta_{x}-\eta _{y}\right) \text{,}  \label{7.20}
\end{equation}
where $\eta_{x}$ and $\eta_{y}$ can be any pair of well-behaved time
coordinates for the points $x$ and $y$, and the star index in the
theta function was used to indicate that the derivative operators
acting on the scalar field which appear in the stress tensor operator
should also act on the theta function. Thus, all the terms in
expression (\ref{7.20}) are either proportional to $\theta
(\eta_{x}-\eta_{y})$, $\delta (\eta _{x}-\eta _{y})$ or $\delta
^{\prime }(\eta _{x}-\eta _{y})$, and, being proportional to a
commutator, expression~(\ref{7.20}) vanishes for spacelike separated
points because of the microcausality condition of the quantum field
theory for the matter fields. Furthermore, since both the divergences
and the counterterms are local (proportional to delta functions or
derivatives of them), the contribution to Eq.~(\ref{7.19}) from the
term $\left( H_\mathrm{ren}\cdot h\right)_{ab}\left( x\right)$ is
causal, \emph{i.e.}, it only depends on the metric perturbations
$h_{cd}(y)$ at any point within the past lightcone of $x$.

Finally, taking into account Eq.~(\ref{7.15b}), Eq.~(\ref{7.19})
becomes
\begin{equation}
G_{ab}^{(1)}\left[ g+h\right] =\kappa \langle \hat{T}_{ab}^{(1)}\left[
g+h\right] \rangle _\mathrm{ren}+\kappa \xi _{ab}\text{,}  \label{7.21}
\end{equation}
where the indices have been lowered using the background metric. It
should be noted that, in contrast to Sec.~\ref{sec2.1}, the tensors
appearing in Eq.~(\ref{7.21}) correspond to perturb the background
tensors with both indices already raised. However, as pointed out in
Appendix~\ref{appA}, the resulting equations in both cases are
equivalent because the unperturbed tensors satisfy the semiclassical
Einstein equation. Therefore, Eq.~(\ref{7.21}) is in complete
agreement with Eq.~(\ref{7.7}), keeping in mind that the finite
contributions of the counterterms, corresponding to $A_{ab}$ and
$B_{ab}$ in Eq.~(\ref{7.3}), have been reabsorbed in the renormalized
expectation value of the stress tensor operator.

After this brief review of the functional approach to the
Einstein-Langevin equation, let us now see how Eq.~(\ref{7.15b}) gives
a result for $\langle \delta \hat{\mathcal{T}}_{0}^{i}\rangle_{\Phi }$
which is equivalent to that obtained in Sec.~\ref{sec3.2}.  To begin
with, it should be pointed out that the ambiguity mentioned in
Appendix~\ref{appA} does not affect the $0i$ component of $\langle
\delta \hat{\mathcal{T}}_{a}^{b}\rangle_{\Phi }$ since both the
background stress tensor and the scalar metric perturbations in the
longitudinal gauge are diagonal. Furthermore, it can be seen that for
a diagonal perturbed metric the $0i$ component of the second term on
the right-hand side of Eq.~(\ref{7.15b}) vanishes. Thus, we can
concentrate on the first term.

Fourier transforming the spatial coordinates as done in
Sec.~\ref{sec3}, the expectation value for the $0i$ component of the
perturbed stress tensor becomes
\begin{equation}
\langle (\hat{T}_{0i}^{(1)}) _{k}[g+h] \rangle (\eta )
=-2\int \sqrt{-g\left( \eta ^{\prime }\right) }d\eta ^{\prime }H_{0icd}(\eta
,\eta ^{\prime };\vec{k})h_{k}^{cd}(\eta ^{\prime }) \text{,}  \label{8.80}
\end{equation}
where the kernel $H_{abcd}(\eta ,\eta ^{\prime };\vec{k})$ corresponds
to the Fourier transform of the two terms on the right-hand side of
Eq.~(\ref{7.12}).  Using the equivalent expression in
Eq.~(\ref{7.20}), the kernel $H_{abcd}(\eta ,\eta ^{\prime };\vec{k})$
is given by the following expression, which already takes into account
that the Fourier transform of the expectation value $\left\langle
  \left[ \hat{t}_{ab}[g;\eta
    ,\vec{x}),\hat{t}_{cd}[g;\eta',\vec{x}')\right] \right\rangle$ is
proportional to a Dirac delta function due to the existing translation
invariance in the spatial coordinates,
\begin{equation}
H_{abcd}(\eta ,\eta ^{\prime };\vec{k})(2\pi )^{3}\delta (\vec{k}
+\vec{k}')=-\frac{i}{4}\left\langle \left[ \left( \hat{t}_{ab}\right)
_{k}[g;\eta ),\left( \hat{t}_{cd}\right) _{k'}[g;\eta')\right]
\right\rangle\,\theta ^{*}(\eta -\eta ^{\prime })
\text{,}  \label{8.80b}
\end{equation}
where the star in the theta function had been introduced earlier to
indicate that the derivatives appearing in $\hat{t}_{ab}$ and
$\hat{t}_{cd}$ should also act on the theta function. Performing a
similar decomposition to that introduced for the noise kernel in
Eq.~(\ref{8.22}), we obtain two non-vanishing contributions to the
expectation value $\left\langle \left[ \hat{t}_{ab}[x;x),
\hat{t}_{cd}[g;x^{\prime })\right]\right\rangle[g]$:
\begin{equation}
\left\langle \left[ \hat{t}_{ab}[g;x),\hat{t}_{cd}[g;x')\right]\right\rangle
=\left\langle \left[ \hat{t}_{ab}[g;x),\hat{t}_{cd}[g;x')\right]
\right\rangle _{\phi \varphi }+\left\langle \left[ \hat{t}_{ab}
[g;x),\hat{t}_{cd}[g;x')\right] \right\rangle _{\varphi \varphi}
\text{,}  \label{8.81}
\end{equation}
where the first contribution is quadratic in the quantum operator
$\hat{\varphi}[g]$ for the inflaton perturbations evolving on the
unperturbed geometry, whereas the second contribution is quartic in
$\hat{\varphi}[g]$.  As already pointed out for the separation of the
noise kernel, the fact that the conservation of the stress tensor,
which is the source of the Einstein equation, is necessary to
guarantee its integrability implies that both contributions to the
expectation value must be separately conserved if we want to discard
one of them keeping the consistency of the Einstein equation at the
order that we are working, which is linear in the metric
perturbations. This is indeed the case as follows from the fact that
both the background homogeneous solution $\phi (\eta )$ and the
operator $\hat{\varphi}[g]$ satisfy the Klein-Gordon equation on the
background spacetime.

If we keep only the first term on the right-hand side of
Eq.~(\ref{8.81}), \emph{i.e.}, if we take $H_{abcd}(x,x')=\left\langle
\left[ \delta\hat{t}_{ab} [g;x),\delta\hat{t}_{cd}[g;x')\right]\right\rangle
\theta^{*}(\eta_x-\eta'_x)$, which corresponds to considering the
contributions to $\hat{t}_{ab}[g]$ that are linear in the inflaton
perturbations and is consistent with the linearization of the inflaton
perturbations that was considered in Secs.~\ref{sec3} and \ref{sec4},
we obtain
\begin{eqnarray}
\langle (\delta \hat{\mathcal{T}}_{0i})_k (\eta)
\rangle_{\Phi } = -i \phi'(\eta) (ik_i) \int_{\eta _{0}}^{\eta
}d\eta' a^2(\eta') \Phi_k(\eta') \left\{4\phi'(\eta')\left\langle
[\hat{\varphi}_k(\eta),\hat{\varphi}'_{-k}(\eta')] \right\rangle
\right. \nonumber \\
\left.
-2m^2a^2(\eta')\phi(\eta') \left\langle
[\hat{\varphi}_k(\eta),\hat{\varphi}_{-k}(\eta')] \right\rangle
\right\} \theta^*(\eta,\eta'),   \label{8.81b}
\end{eqnarray}
where we used the explicit expressions for the components $H_{0i00}$
and $H_{0ijj}$. In this case, there is actually no need for the $*$
prescription in $\theta^*(\eta,\eta')$, which implies that the
derivative acting on $\hat{\varphi}(\eta')$ should also act on the
theta function, since it yields a term proportional to
$[\hat{\varphi}(\eta),\hat{\varphi}(\eta)] \delta(\eta-\eta')$, which
vanishes identically. Integrating by parts the first term in the
integrand and using the Klein-Gordon equation for the background
solution $\phi(\eta)$, given by Eq.~(\ref{8.10}), we finally get
\begin{eqnarray}
\langle (\delta \hat{\mathcal{T}}_{0}^{i})_k(\eta)
\rangle_{\Phi } = i a^{-2}(\eta)\phi'(\eta) (ik_i)\int_{\eta_0}^{\eta }
d\eta' a^2(\eta') \left\{4[\hat{\varphi}_k(\eta),
\hat{\varphi}_{-k}(\eta')] \phi'(\eta')\Phi'_k(\eta')
\right. \nonumber \\
\left.
-2[\hat{\varphi}_k(\eta),\hat{\varphi}_{-k}(\eta')]m^2a^2(\eta')\phi(\eta')
\Phi_k(\eta')\right\},    \label{8.82}
\end{eqnarray}
where the factor $a^{-2}(\eta)$ comes from raising the index $i$ with
the background metric and we have substituted the expectation value
for the commutator of the field operators simply by the commutator
since for a linear theory they are $c$-numbers, whose expectation
value is independent of the state.  This result for the expectation
value of the stress tensor coincides with Eq.~(\ref{8.79}), found in
Sec.~\ref{sec3.2}. It should be noted that the contribution from the
boundary term at $\eta'=\eta$ which results from the integration by
parts vanishes because $[\hat{\varphi}(\eta),
\hat{\varphi}(\eta)]=0$. On the other hand, there is a non-vanishing
contribution from the boundary term at $\eta'=\eta_0$:
\begin{equation}
- 4k_i \phi'(\eta)\phi'(\eta_0) \left(\frac{a(\eta_0)}{a(\eta)}\right)^2
[\hat{\varphi}_k(\eta),\hat{\varphi}_{-k}(\eta_0)] \Phi_k(\eta_0).
\label{b10}
\end{equation}
It might seem that the existence of this term would imply a conflict
between the result for the expectation value of the linearized stress
tensor operator obtained in Sec.~\ref{sec3.2} using the equations of
motion for the quantum operators in the Heisenberg picture and the
result based on the influence functional formalism derived in this
appendix. However, this is not the case. The reason for the apparent
discrepancy is the following. When computing the expectation value of
the stress tensor operator, there are terms proportional to
$\left\langle \hat{\varphi}[g;\eta) \right\rangle$, where the operator
$\hat{\varphi}[g;\eta)$, which satisfies the Klein-Gordon equation,
can be written as a linear combination of a term proportional to
$\hat{\varphi}[g;\eta_0)$ and a term proportional to
$\hat{\varphi}'[g;\eta_0)$. In particular, in Sec.~\ref{sec3} we chose
a state for which both $\langle\hat{\varphi}[g;\eta_0)\rangle$ and
$\langle\hat{\varphi}'[g;\eta_0)\rangle$ vanished. On the other hand,
in the approach based on the influence functional formalism the
operators which naturally determine $\hat{\varphi}[g;\eta)$ in terms
of the initial state are $\hat{\varphi}[g;\eta_0)$ and its conjugate
momentum $\hat{\pi}[g;\eta_0)$. Since the coupling between the metric
perturbations and the inflaton perturbations involves terms
proportional to the time derivative of the inflaton perturbations,
$\hat{\pi}[g;\eta_0)$ will differ from $\hat{\varphi}'[g;\eta_0)$ by a
term proportional to the metric perturbations at the initial
time. This is precisely the origin of the term in
expression~(\ref{b10}). Thus, the apparent discrepancy is just a
consequence of the fact that in the influence functional approach it
has been implicitly assumed that the initial state has vanishing
$\left\langle \hat{\pi}[g;\eta_0) \right\rangle$ rather than vanishing
$\left\langle \hat{\varphi}'[g;\eta_0) \right\rangle$.

It is important to stress that the expression in Eq.~(\ref{8.82}) for
the expectation value of the stress tensor operator needs no
renormalization. This fact can be easily understood because we are
dealing with the linearized theory. Therefore, the terms involved in
the computation of the expectation value of the stress tensor operator
are proportional to $\left\langle \hat{\varphi}[g+h;x] \right\rangle$,
whereas the divergences that arise in an exact treatment (without
linearizing with respect to the scalar field) are a consequence of
taking the coincidence limit $x' \rightarrow x$ in terms involving
products of the field operator, \emph{i.e.}, proportional to
$\left\langle\hat{\varphi}[g+h;x]\hat{\varphi}[g+h;x']\right\rangle$.
Alternatively, when considering Eq.~(\ref{7.14}) together with
Eq.~(\ref{7.20}), the need for renormalization can be understood as
follows. The expectation value of the commutator is finite as long as
one restricts to $x \neq x'$, but it diverges when one considers the
coincidence limit. Nevertheless, it is still meaningful as a
distribution. In this context, the divergences arise because the
product of distributions in Eq.~(\ref{7.20}) is ill defined in general
although each factor is well defined as a distribution; see
Ref.~\cite{roura99a} for a detailed discussion on this point. In fact,
the terms in Eq.~(\ref{7.20}) involve terms proportional to the
imaginary part of the product of two Feynman propagators
$G_F(x,x')G_F(x,x')$ \cite{campos94,campos96,martin00}. Working in
Fourier space for the spatial variables, this product becomes $\int
d^3q {G}_F(\eta,\eta';\vec{k}-\vec{q}) {G}_F(\eta,\eta';\vec{q})$,
which exhibits an ultraviolet divergence when performing the integral
$\int d^3q$ over all possible momenta. On the other hand, when
linearizing with respect to the scalar field, the Fourier transformed
version of the terms in Eq.~(\ref{7.20}) is simply proportional to
${G}_F(\eta,\eta';\vec{k})$, with no integral over momenta and, hence,
no ultraviolet divergence.

\section{Initial conditions}

\label{appC}

In this appendix we will explain why, strictly speaking, a homogeneous
solution $\Phi_k^\mathrm{(h)}(\eta)$ with some particular initial
conditions must be added to the purely inhomogeneous solution
$\Phi_k^\mathrm{(i)}(\eta)$ when solving the Einstein-Langevin
equation as done in Secs.~\ref{sec3} and \ref{sec4}. The situation is
completely analogous when solving the quantum version of the
linearized Einstein equation as in Sec.~\ref{sec3.3}.

It is well known that the Bianchi identity guarantees the
integrability of the Einstein equation provided that the stress tensor
of the matter sources is covariantly conserved. Let us, however,
discuss this point in some more detail. The ten components of the
Einstein equation for a globally hyperbolic spacetime, which can be
foliated with a set of Cauchy hypersurfaces, can be formulated as an
initial value problem with time corresponding to some continuous
variable labeling the Cauchy hypersurfaces. In particular for the
cosmological problem that we are considering we can choose the
homogeneous spatial sections labeled by the conformal time $\eta$ as
the set of Cauchy hypersurfaces. The four temporal components of the
Einstein equation can then be regarded as a set of dynamical
constraints at any given instant of time. Thus, the integrability of
the Einstein equation as an initial value problem can be understood in
the following way: using the Bianchi identity and the conservation of
the matter sources, the constraints can be shown to hold at any time
provided that the spatial components of the Einstein equation are
satisfied for all times and the four constraint equations are
fulfilled on the Cauchy hypersurface corresponding to some initial
time \cite{wald84}. Obviously, the previous discussion can be extended
to the case of the Einstein-Langevin equation since the stochastic
source is also covariantly conserved.

Let us recall the temporal components of the Einstein-Langevin
equation for scalar metric perturbations after Fourier transforming
with respect to the spatial coordinates:
%\begin{eqnarray}
%\left[k^2-\left(\mathcal{H}'+2\mathcal{H}^2\right)\right] \Phi_k
%- 3\mathcal{H}\Phi'_k = \frac{\kappa }{2}a^{2} \left(
%\left\langle \left( \delta \hat{\mathcal{T}}_{0}^{0}\right)_k\right\rangle
%_{\Phi } + \left(\xi_{0}^{0}\right)_{k} \right), \label{c1} \\
%ik_{i}(\Phi'_{k} + \mathcal{H}\Phi_{k}) = \frac{\kappa}{2} a^{2} \left(
%\left\langle \left( \delta \hat{\mathcal{T}}_{0}^{i}\right)_k\right\rangle
%_{\Phi } + \left(\xi_{0}^{i}\right)_{k} \right). \label{c2}
%\end{eqnarray}
\begin{eqnarray}
\left( k^2 + 3\mathcal{H}^2 \right) \Phi_k
+ 3\mathcal{H}\Phi'_k = \frac{\kappa }{2}a^{2} \left(
\langle ( \delta \hat{\mathcal{T}}_{0}^{0})_k\rangle
_{\Phi } + \left(\xi_{0}^{0}\right)_{k} \right), \label{c1} \\
ik_{i}(\Phi'_{k} + \mathcal{H}\Phi_{k}) = \frac{\kappa}{2} a^{2} \left(
\langle ( \delta \hat{\mathcal{T}}_{0}^{i})_k\rangle
_{\Phi } + \left(\xi_{0}^{i}\right)_{k} \right). \label{c2}
\end{eqnarray}
In Secs.~\ref{sec3} and \ref{sec4} the constraint equation (\ref{c2})
was solved to find $\Phi_k(\eta)$. However, one should make sure that
the remaining components of the Einstein-Langevin equation are also
satisfied. According to the discussion in the previous paragraph, to
make sure that this is indeed the case it is sufficient to demand that
the Eq.~(\ref{c1}) holds at the initial time $\eta_0$ for every
$\vec{k}$. The solution of Eq.~(\ref{c2}) can always be written as
$\Phi_k(\eta) = \Phi_k^\mathrm{(h)}(\eta) +
\Phi_k^\mathrm{(i)}(\eta)$, where $\Phi_k^\mathrm{(i)}(\eta)$ is a
solution of the inhomogeneous equation which vanishes at the initial
time $\eta_0$ and $\Phi_k^\mathrm{(h)}(\eta)$ is a solution of the
homogenous equation which is completely determined by specifying its
value at $\eta_0$. Imposing Eq.~(\ref{c1}) at $\eta_0$ and using
Eq.~(\ref{c2}) evaluated at $\eta_0$ in order to substitute
$\Phi'_k(\eta_0)$, one obtains the following result for
$\Phi_k^\mathrm{(h)}(\eta_0)$:
%\begin{equation}
%\Phi_k^\mathrm{(h)}(\eta_0)=\frac{\kappa}{2}a^2(\eta_0)
%\left(k^2 +\mathcal{H}^2(\eta_0) -\mathcal{H}'(\eta_0)
%- \frac{\kappa}{2} \left(\phi'(\eta_0)\right)^2 \right)^{-1}
%\left[\left(\xi_{0}^{0}\right)_{k} (\eta_0) +
%\frac{3 \mathcal{H}(\eta_0)}{ik_i} \left(\xi_{0}^{i}\right)_{k} (\eta_0)
%\right], \label{c3}
%\end{equation}
\begin{eqnarray}
\Phi_k^\mathrm{(h)}(\eta_0) &=& \frac{\kappa}{2}a^2(\eta_0)
\left(k^2 - \frac{\kappa}{2} \left(\phi'(\eta_0)\right)^2 \right)^{-1}
\left[\left(\xi_{0}^{0}\right)_{k} (\eta_0) -
\frac{3 \mathcal{H}(\eta_0)}{ik_i} \left(\xi_{0}^{i}\right)_{k} (\eta_0)
\right] \nonumber \\
&=& \frac{\kappa}{2}a^2(\eta_0)
\left(k^2 + \mathcal{H}'(\eta_0) - \mathcal{H}^2(\eta_0) \right)^{-1}
\left[\left(\xi_{0}^{0}\right)_{k} (\eta_0) -
\frac{3 \mathcal{H}(\eta_0)}{ik_i} \left(\xi_{0}^{i}\right)_{k} (\eta_0)
\right],  \label{c3}
\end{eqnarray}
where we took into account that $\langle
(\delta\hat{\mathcal{T}}_{0}^{i})_k (\eta_0)\rangle_{\Phi }$ vanishes,
as can be immediately seen from Eqs.~(\ref{8.79}), or (\ref{8.81b}),
because the limits of integration coincide. We also used the fact that
$\langle (\delta\hat{\mathcal{T}}_{0}^{0})_k (\eta_0)\rangle_{\Phi } =
a^{-2}(\eta_0) [\phi'(\eta_0)]^2 \Phi_k (\eta_0)$: in this case the
second term on the right-hand side of Eq.~(\ref{7.14}) vanishes for
the same reason as with $\langle (\delta\hat{\mathcal{T}}_{0}^{i})_k
(\eta_0)\rangle_{\Phi }$, but there is a non-vanishing contribution
from the last term in Eq.~(\ref{7.14}), which corresponds to the first
term on the right-hand side of Eq.~(\ref{8.49}).

%******************************************************

Since Eq.~(\ref{c2}) is a first order integro-differential equation,
the result for $\Phi_k^\mathrm{(h)}(\eta_0)$ in Eq.~(\ref{c3})
completely determines the homogeneous solution
$\Phi_k^\mathrm{(h)}(\eta)$.  The situation will be completely
analogous when linearizing and quantizing both the metric
perturbations and the inflaton perturbations, as done in
Sec.~\ref{sec3.3}, with the quantum operator for the metric
perturbations $\hat{\Phi}(x)$ replacing the stochastic scalar field
$\Phi(x)$ and the operator $\delta \hat{t}_a^b$ instead of the
stochastic source $\xi_a^b(x)$. Hence, the argument concerning the
equivalence between the quantum correlation function for the metric
perturbations and the stochastic correlation function can be
straightforwardly extended, following the same line of reasoning as in
Sec.~\ref{sec3.3}, to the case in which the contribution from the
homogeneous solution is also taken into account.

Nevertheless, in Secs.~\ref{sec3} and \ref{sec4} this homogeneous
solution was not considered when giving the final result for the
correlation function of the metric perturbations.
%Let us, therefore, discuss those situations in which the contribution
%from the homogeneous solution can be safely neglected.
%
%A case in which the contribution from the homogeneous solution
%vanishes is that of asymptotically past initial conditions,
%\emph{i.e.}, when $\eta_0 \rightarrow -\infty$ and it is implicitly
%assumed that the gravitational interaction is adiabatically switched
%on. Hence, $\Phi_k^\mathrm{(h)}(\eta_0)$ vanishes because we should let
%$\kappa \rightarrow 0$ in Eq.~(\ref{c3}) as $\eta_0 \rightarrow
%-\infty$. Note that it is not consistent to switch on the interaction
%in a finite time even if it is done smoothly because the source of the
%Einstein-Langevin equation (the gravitational constant times the sum
%of the stochastic source plus the expectation value of the stress
%tensor operator) would no longer be conserved.
%
Therefore, we end this appendix arguing why it is justified to neglect
the contribution from the homogeneous solution when computing the
correlation functions for scalar metric perturbations at large scales
in the context of cosmological inflationary models. In other words,
the contribution from the first three terms on the right-hand side of
Eq.~(\ref{8.72b}) is much smaller than the contribution from the
fourth term when considering a situation similar to that addressed in
Sec.~\ref{sec4}. This can be qualitatively understood in the following
way. Since Eq.~(\ref{c2}) is a linear first order differential
equation, the solution of the homogeneous equation,
$\Phi_k^\mathrm{(h)}(\eta)$, will be proportional to the expression
for $\Phi_k^\mathrm{(h)}(\eta)$ given by Eq.~(\ref{c3}). Thus, the
first term on the right-hand side of Eq.~(\ref{8.72b}) is proportional
to the correlation functions for the stochastic source at the time
$\eta_0$, and the second and third terms are proportional to the
correlation functions at different times: $\eta_0$ and a time $\eta'$
which is integrated from $\eta_0$ to $\eta_1$ or $\eta_2$ [see
Eq.~(\ref{8.72})]. The value of the noise kernel is small when one or
both the two arguments are $\eta_0$ provided that $\eta_0$ is negative
enough so that the scales of interest were well inside the horizon at
that time. This is in contrast to the contribution from the last term
in Eq.~(\ref{8.72b}) when the relevant scales are well outside the
horizon at $\eta_1$ and $\eta_2$, since the two arguments of the noise
kernel in that term are integrated from $\eta_0$ to $\eta_1$ or
$\eta_2$. Hence, the reason for neglecting the first three terms on
the right-hand side of Eq.~(\ref{8.72b}) in this context is actually
rather similar to the reason for the weak dependence on $\eta_0$ of
the result obtained in Sec.~\ref{sec4} when the scale $k$ is well
outside the horizon at $\eta_1$ and $\eta_2$, and $\eta_0$ is negative
enough so that $k$ is well inside the horizon at that time.

The previous argument can be made more precise if we concentrate on
the particular model considered in Sec.~\ref{sec4}. In that case, if
we neglect the non-local term corresponding to $\langle \delta
\hat{\mathcal{T}}_{0}^{i} \rangle_{\Phi }$, as done in
Sec.~\ref{sec4}, the expression for the homogenous solution is
\begin{equation}
\Phi_k^\mathrm{(h)}(\eta)=\frac{\kappa}{2}\frac{a^3(\eta_0)}{a(\eta)}
\left(k^2 + \mathcal{H}'(\eta_0) - \mathcal{H}^2(\eta_0) \right)^{-1}
\left[\left(\xi_{0}^{0}\right)_{k} (\eta_0) -
\frac{3 \mathcal{H}(\eta_0)}{ik_i} \left(\xi_{0}^{i}\right)_{k} (\eta_0)
\right]. \label{c4}
\end{equation}
The contribution form the first three terms on the right-hand side of
Eq.~(\ref{8.72b}) can then be explicitly computed and compared to the
last term, taking into account that $k\eta_1, k\eta_1\ll 1$ and
$k\eta_0 \gg 1$. In particular, the first term on the right-hand side
of Eq.~(\ref{8.72b}) is proportional to $(\kappa/2)(m/m_p)^2(2\pi)^3
k^{-3} \delta(\vec{k}-\vec{k}') a^2(\eta_0)/a(\eta_1)a(\eta_2)$ and a
sum of terms of order $1$, $(1/k\eta_0)^2$ and
$(m/H)(1/k\eta_0)^2$. The factor $a^2(\eta_0)/a(\eta_1)a(\eta_2)$,
which is of order $k\eta_1k\eta_2/(k\eta_0)^2$, as well as
$(1/k\eta_0)^2$ and $m/H$ are much smaller than $1$. It is thus clear
that those contributions can be safely neglected as compared to the
last term, which was found to be of order $(\kappa/2)(m/m_p)^2(2\pi)^3
k^{-3} \delta(\vec{k}-\vec{k}')$ in Sec.~\ref{sec4}.

Similarly, the second and third terms on the right-hand side of
Eq.~(\ref{8.72b}) are proportional to $(\kappa/2)(m/m_p)^2(2\pi)^3
k^{-3} \delta(\vec{k}-\vec{k}') a^2(\eta_0)/a(\eta_1)a(\eta_2)$ and a
sum of terms of order $1$ and $1/k\eta_0$. Therefore, they can also be
neglected as compared to the last term.

\section{Alternative proof of the equivalence between stochastic and
quantum correlation functions}

\label{appD}

In this appendix we provide an alternative proof of the equivalence between stochastic and quantum correlation functions whose key step is to show that the Einstein-Langevin equation for linearized cosmological perturbations implies Eq.~(6.48) of Ref.~\cite{mukhanov92}.

Let us consider the Einstein-Langevin equation for scalar metric perturbations when one also linearizes with respect to the inflaton field, whose different components are given by  Eqs.~(\ref{8.70})-(\ref{8.71b}). We will take $\Psi = \Phi$, as justified by the discussion before Eq.~(\ref{8.71c}), and work in Fourier space for the spatial coordinates. Next, we add  Eq.~(\ref{8.70}), the $i=j$ component of Eq.~(\ref{8.71b}) and Eq.~(\ref{8.71}) multiplied by $F \equiv 2 m^2 a^2 (\phi/\phi') (k_i/k^2)$, which leads to the following result: 
\begin{equation}
\Phi''_k + 2 \left( \mathcal{H} - \frac{\phi''}{\phi'} \right) \Phi'_k + k^2 \Phi_k
+ 2 \left( \mathcal{H}' - \mathcal{H} \frac{\phi''}{\phi'} \right) \Phi_k
= \frac{\kappa }{2}a^{2}\left[ \langle \delta \hat{\mathcal{T}}_{0}^{0}
\rangle _{\Phi } + \langle \delta \hat{\mathcal{T}}_{i}^{i}
\rangle _{\Phi } + F \langle \delta \hat{\mathcal{T}}_{0}^{i}
\rangle _{\Phi } + \xi _{0}^{0} + \xi _{i}^{i} + F \xi _{0}^{i} \right]_k
\label{d1} ,
\end{equation}
with no summation over the repeated $i$ indices. In deriving Eq.~(\ref{d1}) we made use of the following two relations
\begin{eqnarray}
6 \mathcal{H} + 2m^2 a^2 \frac{\phi}{\phi'} = \frac{2}{\phi'} \left( 3 \mathcal{H} \phi'
+ m^2 a^2 \phi  \right) &=& 2 \mathcal{H} - 2 \frac{\phi''}{\phi'}
\label{d2}, \\
2 \mathcal{H}' + 2 \mathcal{H} \left( 2 \mathcal{H}
+  m^2 a^2 \frac{\phi}{\phi'} \right)
&=& 2 \mathcal{H}' - 2 \mathcal{H} \frac{\phi''}{\phi'}
\label{d3},
\end{eqnarray}
which follow from the Klein-Gordon equation~(\ref{8.10}) for the background field $\phi$. The final step is to show that the right-hand side of Eq.~(\ref{d1}) vanishes. In order to do so, it is convenient to consider first the Fourier-transformed version of Eqs.~(\ref{8.49})-(\ref{8.51}) for the linearized stress tensor. It is then straightforward to show that 
\begin{equation}
\left(\delta \mathcal{T}_{0}^{0}\right)_k + \left(\delta \mathcal{T}_{i}^{i}\right)_k
+ F \left(\delta \mathcal{T}_{0}^{i}\right)_k = 0
\label{d4} ,
\end{equation}
with no summation over the repeated $i$ indices. The same conclusion applies when $\varphi$ is promoted to a Heisenberg operator $\hat{\varphi}$, which implies that the first three terms on the right-hand side of Eq.~(\ref{d1}) cancel out. On the other hand, since $\left(\xi_{0}^{0} + \xi_{i}^{i} + F \xi_{0}^{i}\right)_k$ is a Gaussian stochastic process with vanishing mean, in order to prove that it vanishes it is sufficient to see that $\left \langle \left(\xi_{0}^{0} + \xi_{i}^{i} + F \xi_{0}^{i}\right)_k \! (\eta) \, \left(\xi_{\mu}^{\nu}\right)_{k'} (\eta') \right \rangle_\xi$, which is proportional to $\left \langle \left\{\left(\delta\hat{t}_{0}^{0} + \delta\hat{t}_{i}^{i} + F \delta\hat{t}_{0}^{i}\right)_k \! (\eta), \, \left(\delta\hat{t}_{\mu}^{\nu}\right)_{k'} (\eta') \right\} \right \rangle$, vanishes. Indeed, taking $\Phi = 0$ in Eq.~(\ref{d4}) and promoting  $\varphi$ to a Heisenberg operator, it follows that $\left(\delta\hat{t}_{0}^{0} + \delta\hat{t}_{i}^{i} + F \delta\hat{t}_{0}^{i}\right)_k = 0$. Thus, the right-hand side of Eq.~(\ref{d1}) vanishes and one is left with
\begin{equation}
\Phi'' + 2 \left( \mathcal{H} - \frac{\phi''}{\phi'} \right) \Phi' - \nabla^2 \Phi
+ 2 \left( \mathcal{H}' - \mathcal{H} \frac{\phi''}{\phi'} \right) \Phi = 0
\label{d5} ,
\end{equation}
which coincides with Eq.~(6.48) in Ref.~\cite{mukhanov92}. Several remarks about Eq.~(\ref{d5}) are in order. First, the non-local terms associated with $\langle \delta \hat{\mathcal{T}}_{a}^{b}\rangle _{\Phi }$ are not present so that, when working in Fourier space for the spatial coordinates, one is left with an ordinary differential equation rather than an integro-differetnial one. Second, the equation exhibits no dependence on the stochastic source. However, the solutions of the Einstein-Langevin equation should also satisfy the constraint equations at the initial time in addition to Eq.~(\ref{d5}). According to the results in Appendix~\ref{appC}, this implies a dependence on the stochastic source for the initial conditions $\Phi_k (\eta_0)$ and $\Phi_k' (\eta_0)$, which will  involve a linear combination of terms linearly proportional to the stochastic source [as given by Eq.~(\ref{c3}) and an analogous result for $\Phi_k' (\eta_0)$ that can be obtained by substituting Eq.~(\ref{c3}) into Eq.~(\ref{c2})]. The solution of the linearized Einstein-Langevin equation can then be written as $\Phi_k (\eta) = u_1(\eta) \Phi_k (\xi_\mu^\nu(\eta_0);\eta_0) + u_2(\eta) \Phi_k' (\xi_\mu^\nu (\eta_0);\eta_0)$, where $u_1(\eta)$ and $u_2(\eta)$ are solutions of Eq.~(\ref{d5}) with initial conditions $u_1(\eta_0)=1,\, u_1'(\eta_0)=0$ and $u_2(\eta_0)=0,\, u_2'(\eta_0)=1$. Such a dependence of the initial conditions on the stochastic source at the initial time is responsible for the non-trivial part of the stochastic correlation functions of $\Phi$ at later times.

If one quantizes both the linearized metric perturbation and the inflaton field, so that $\Phi$ and $\varphi$ are promoted to Heisenberg operators in Eqs.~(\ref{8.53})-(\ref{8.55}), one can easily conclude (proceeding analogously to the previous paragraph) that the operator $\hat{\Phi}$ for the linearized metric perturbations also satisfies Eq.~(\ref{d5}).
Furthermore, the constraints at the initial time give the same results for $\hat{\Phi}_k (\eta_0)$ and $\hat{\Phi}_k' (\eta_0)$ as in the stochastic case but with $\delta\hat{t}_\mu^\nu$ in place of $\xi_\mu^\nu$, so that the solutions of the quantum version of Eqs.~(\ref{8.53})-(\ref{8.55}) can be written as $\hat{\Phi}_k (\eta) = u_1(\eta) \hat{\Phi}_k (\delta\hat{t}_\mu^\nu(\eta_0);\eta_0) + u_2(\eta) \hat{\Phi}_k' (\delta\hat{t}_\mu^\nu(\eta_0),\eta_0)$. Taking into account that $\left\langle \xi _\mu^\nu \left[ g;x\right) \xi_\rho^\sigma \left[ g;y\right)\right\rangle _{\xi } = (1/2) \left\langle
\left\{ \delta\hat{t}_\mu^\nu \left[ g;x\right), \delta\hat{t}_\rho^\sigma \left[ g;y\right) \right\}
\right\rangle$, it is straightforward to see that the result for symmetritzed two-point quantum correlation function $(1/2) \langle \{ \hat{\Phi}_k(\eta _{1}),\hat{\Phi}_{k'}(\eta _{2})\} \rangle$ is equivalent to that for the stochastic correlation function $\langle \Phi_k(\eta _{1}) \Phi_{k'}(\eta _{2}) \rangle_\xi$. This constitutes an alternative proof to that provided in Sec.~\ref{sec3.3} of the equivalence between quantum and stochastic correlation functions for linearized cosmological perturbations. 

We close this appendix by discussing a recent claim that there is a discrepancy between stochastic gravity and the standard treatment for superhorizon modes. More specifically, in Ref.~\cite{urakawa08a} the Einstein-Langevin equation for linearized cosmological perturbations was solved using certain approximations and the correlation function for the gauge invariant variable $\zeta$ was computed (this variable corresponds to the curvature perturbation in the uniform density gauge \cite{wands00}, which coincides with the comoving gauge for modes outside the horizon \cite{lyth03}). It was found that for modes outside the horizon and for a sufficiently large number of e-folds the correlation function was not constant in time, contrary to the standard result. This would be in conflict with the equivalence for linear perturbations that we have shown to hold in general. However, one can provide an exact argument which shows that the linearized Einstein-Langevin equation actually implies that for modes outside the horizon the correlation function of $\zeta$ remains constant in time. This means that the result in Ref.~\cite{urakawa08a} seems to imply a problem with some of their approximations rather than a shortcoming of stochastic gravity for that regime as concluded there.

The exact argument is the following. One starts with the expression for $\zeta$ in terms of $\Phi$ \cite{lyth85,mukhanov92,lyth03}:
\begin{equation}
\zeta = \frac{2}{3(1+w)} ( \Phi + \mathcal{H}^{-1} \Phi' ) + \Phi
\label{d6} ,
\end{equation}
with $w=p/\rho$, where the relation between the background density $\rho$ and pressure $p$ and the background field $\phi$ is given by Eqs.~(\ref{8.5})-(\ref{8.6}) and the text after them. Differentiating Eq.~(\ref{d6}) with respect to the conformal time and multiplying by $(3/2)  \mathcal{H} (1+w)$, one obtains
\begin{equation}
\frac{3(1+w)}{2} \mathcal{H} \zeta' = \Phi'' + 2 \left( \mathcal{H}
- \frac{\phi''}{\phi'} \right) \Phi' + 2 \left( \mathcal{H}'
- \mathcal{H} \frac{\phi''}{\phi'} \right) \Phi
\label{d7} ,
\end{equation}
where we made use of the following relations, which can be derived from Eqs.~(\ref{8.8})-(\ref{8.9}),
\begin{eqnarray}
&& 1+w=\frac{\rho + p}{\rho} = \frac{\kappa (\phi')^2}{3 \mathcal{H}^2}
=  \frac{2(\mathcal{H}^2 - \mathcal{H}')}{3 \mathcal{H}^2}
\label{d8} ,\\
&& \frac{d}{d \eta} \ln (1+ w) = 2 \frac{\phi''}{\phi'}
- 2 \frac{\mathcal{H}'}{\mathcal{H}}
\label{d9} .
\end{eqnarray}
Note that the right hand-side of Eq.~(\ref{d7}) coincides with the left-hand side of Eq.~(\ref{d5}) except for the $\nabla^2 \Phi$ term. This means that for modes outside the horizon, for which $\nabla^2 \Phi$ can be neglected, the right-hand side of Eq.~(\ref{d7}) vanishes and the mode $\zeta_k$ remains constant in time.

%\pagebreak[4]

%\bibliography{cosmol11}

\end{document}